\documentclass[12pt,letterpaper]{article}

\usepackage{geometry}
\usepackage{amsmath}
\usepackage{graphicx}
\usepackage{enumerate}
\usepackage{natbib}
\usepackage{url} 
\usepackage{amssymb}
\usepackage{dsfont} 

\usepackage{xr}
\usepackage{amsthm} 
\theoremstyle{remark}  
\newtheorem{theorem}{Theorem} 
\theoremstyle{remark}  
\usepackage{arydshln} 
\usepackage{caption} 
\usepackage[table]{xcolor}
\definecolor{light-gray}{gray}{0.7}
\definecolor{light-int-gray}{gray}{0.77}
\definecolor{light-light-gray}{gray}{0.9}

\pdfoutput=1 

\usepackage[utf8]{inputenc}
\usepackage{cite}
\usepackage{nameref,hyperref}
\usepackage[right]{lineno}
\usepackage{microtype}
\DisableLigatures[f]{encoding = *, family = * }

\usepackage{changepage}
\usepackage[aboveskip=1pt,labelfont=bf,labelsep=period,singlelinecheck=off]{caption}
\makeatletter
\renewcommand{\@biblabel}[1]{\quad#1.}
\makeatother

\usepackage{lastpage,fancyhdr,graphicx}
\usepackage{epstopdf}
\usepackage{natbib}
\usepackage{amsmath}
\fancyheadoffset[L]{2.25in}
\fancyfootoffset[L]{2.25in}

\usepackage{color}

\definecolor{Gray}{gray}{.25}

\usepackage{graphicx}

\usepackage{wrapfig}
\usepackage[pscoord]{eso-pic}
\usepackage[fulladjust]{marginnote}
\reversemarginpar

\usepackage{algorithm} 
\usepackage[noend]{algpseudocode}
\usepackage{enumitem}
\makeatletter
\def\BState{\State\hskip-\ALG@thistlm}
\makeatother

\begin{document}
\vspace*{0.35in}
\begin{flushleft}
\large{\hspace{3.95cm}\textbf{Conformal Prediction Bands}}

\large{\hspace{3.35cm}\textbf{for Multivariate Functional Data}}
\newline
\\
\hspace{0.75cm} Jacopo Diquigiovanni\textsuperscript{1,*},
  Matteo Fontana\textsuperscript{2,3},
  Simone Vantini\textsuperscript{2}
\\
\bigskip
{\fontsize{10}{10}\selectfont \bf{1} Department of Statistical Sciences, University of Padova, Italy}
\\
{\fontsize{10}{10}\selectfont \bf{2} MOX - Department of Mathematics, Politecnico di Milano, Italy}
\\
{\fontsize{10}{10}\selectfont \bf{3} now at Joint Research Centre - European Commission, Ispra (VA), Italy}
\\
\bigskip
{\fontsize{10}{10}\selectfont* jacopo.diquigiovanni@phd.unipd.it}

\end{flushleft}

\section*{Abstract}
Motivated by the pressing request of methods able to create prediction sets in a general regression framework for a multivariate functional response  and pushed by new methodological advancements in non-parametric prediction for functional data, we propose a set of conformal predictors that produce finite-sample either valid or exact multivariate simultaneous prediction bands under the mild assumption of exchangeable regression pairs. The fact that the prediction bands can be built around any regression estimator and that can be easily found in closed form yields a very widely usable method, which is fairly straightforward to implement. In addition, we first introduce and then describe a specific conformal predictor that guarantees an asymptotic result in terms of efficiency and inducing prediction bands able to modulate their width based on the local behavior and magnitude of the functional data. The method is investigated and analyzed through a simulation study and a real-world application in the field of urban mobility.

\noindent%
{\it Keywords:}  Functional data; Conformal Prediction; Prediction band; Exact prediction set; Distribution-free prediction set; Finite-sample prediction set

\section{Introduction}

Functional Data Analysis \citep[FDA,][]{ramsay_functional_2005} is now a fairly established, but still very ebullient field of statistics whose goal is to develop theory and methods to treat datasets composed of smooth functions.
Since the first seminal paper by Jim O. Ramsay \citep{ramsay_when_1982}, many standard multivariate tools have been translated to the functional realm: among those Functional Principal Component Analysis \citep[Chapter 10]{ramsay_functional_2005}, Functional Linear Regression  \citep[Chapter 12]{ramsay_functional_2005} and functional boxplots \citep{sun_functional_2011}, just to give a very partial and non-exhaustive list.

A crucial challenge in FDA is the issue of uncertainty quantification in prediction. Intuitively, we are interested in creating prediction sets, namely subsets of the sample space including a new functional observation with a certain nominal confidence level $1-\alpha$.
Only very recent works in FDA provide some knowledge into this theoretical (but yet full of applied repercussions) issue, all of them focusing on the univariate setting (i.e. a framework in which the functional observation consists of a single real-valued function defined over a domain).
These approaches can be classified in three groups: the first one consists of works principally based on parametric bootstrapping techniques \citep[e.g.,][]{degras_simultaneous_2011, cao_simultaneous_2012}, the second one is characterized by the application of dimensionality reduction techniques to manage the naturally infinite dimensionality \citep[e.g.,][]{hyndman_robust_2007,antoniadis_prediction_2016}. These first two groups carry obvious drawbacks since they are either  based on not easily provable distributional assumptions and/or on asymptotic results. In addition, the first class of approaches is computationally demanding, whereas the second one relies on the approximations induced by basis projection.
The third group is based on a novel approach to forecasting in the framework of Conformal Prediction (CP) \citep{diquigiovanni2021importance}. This approach is able to output either exact or valid prediction bands under minimal distributional assumptions and in an efficient way, thus bypassing the methodological shortcomings identified in the previous literature. However, this is done in the setting of univariate i.i.d. functional data.
The objective of the present work is to build from that contribution by extending the method to multivariate functional data and to a regressive framework.

Formally, we will consider independent and identically distributed regression pairs $\boldsymbol{Z}_1,\dots,\boldsymbol{Z}_n \sim P$, with $\boldsymbol{Z}_i=(\boldsymbol{X}_i,\boldsymbol{Y}_i)$ consisting of a multivariate functional response variable $\boldsymbol{Y}_i$ and a set of (not necessarily scalar) covariates $\boldsymbol{X}_i$ $\forall i=1,\dots,n$. Let $\boldsymbol{Y}_i=(Y_{i1},Y_{i2},\dots,Y_{ip})$ be a multivariate random function such that its $j$-th component $Y_{ij}$ ($j=1,\dots,p$) is a random function taking values in  $L^{\infty}(\mathcal{T}_j)$, which is the family of limited functions $y: \mathcal{T}_j \rightarrow \mathbb{R}$  with $\mathcal{T}_j$ closed and bounded subset of $\mathbb{R}^{d_j}$, $d_j \in \mathbb{N}_{>0}$. For the sake of brevity, later in the discussion we will indicate  the space $L^{\infty}(\mathcal{T}_1) \times \dots \times L^{\infty}(\mathcal{T}_p)$ in which $\boldsymbol{Y}_i$ takes values as $\prod_{j=1}^p L^{\infty}(\mathcal{T}_j)$. Note that the framework considered is extremely wide since both the domain $\mathcal{T}_j$ and the image of $Y_{ij}$ are allowed to be very different when $j$ varies. $\boldsymbol{X}_i=(X_{i1},X_{i2},\dots,X_{ip})$ is a set of covariates such that its element related to the $j$-th component $X_{ij}$ (which is a set of covariates itself) belongs to a measurable space and can be very general: for example, $X_{ij}$ can be the usual vector of predictors, or it can be a set of functional covariates allowing for a functional-on-functional regression model, or it can contain both scalar and functional predictors. Let $\mu^j(x_{ij})=\mathbb{E}(Y_{ij}| X_{ij}=x_{ij})$ denote the regression function for the $j$-th component of the $i$-th observation, and consistently with this notation let us define the scalar value $[\mu^j(x_{ij})](t)=\mathbb{E}(Y_{ij}(t)| X_{ij}=x_{ij})$.

The aim of the article is to build a procedure able to output exact (or at least valid) multivariate functional prediction bands under no assumptions on $P$ and $\mu^1(\cdot),\dots,\mu^p(\cdot)$ other than i.i.d. regression pairs. A multivariate functional prediction band is a specific kind of prediction set that can be defined, consistently with the well-known definition of univariate functional prediction band \citep{lopez2009concept,degras2017simultaneous}, as
\begin{equation*}
\left\{\boldsymbol{y}=(y_1,\dots,y_p) \in \prod_{j=1}^p L^{\infty}(\mathcal{T}_j): y_j(t) \in B_{j}(t), \quad \forall j \in 1,\dots,p, \quad \forall t \in \mathcal{T}_j,\right\}
\end{equation*}
with $B_{j}(t)$ interval $\forall j,t$. Prediction bands are so relevant in the functional set prediction framework due to their conceptual simplicity and because they  can be plotted in parallel coordinates  \citep{inselberg1985plane}. A detailed discussion of the topic is provided by \citet{diquigiovanni2021importance}. For the sake of simplicity, later in the discussion the term \textit{prediction band} will be used to indicate a multivariate functional prediction band, unless otherwise specified.

The terms \textit{valid prediction set} and \textit{exact prediction set} are instead used to indicate the coverage ensured by a prediction set.

\paragraph{Valid prediction set} Consistently with the notation of \citet{lei2018distribution}, a valid prediction set for  $\boldsymbol{Z}_{n+1}=(\boldsymbol{X}_{n+1}, \boldsymbol{Y}_{n+1})$ - which is independent from and identically distributed to $\boldsymbol{Z}_{1},\dots, \boldsymbol{Z}_{n}$ - is the set $\mathcal{C}_{n,1-\alpha}$ based on $\boldsymbol{Z}_{1},\dots, \boldsymbol{Z}_{n}$ such that
\begin{equation}
\mathbb{P}\left(\boldsymbol{Y}_{n+1} \in \mathcal{C}_{n,1-\alpha}\left(\boldsymbol{X}_{n+1}\right)\right) \geq 1-\alpha
\label{eq::valid_pred_set}
\end{equation}
for any significance level $\alpha \in (0,1)$, with $\mathcal{C}_{n,1-\alpha}\left(\boldsymbol{x}\right)=\{ \boldsymbol{y} \in \prod_{j=1}^p L^{\infty}(\mathcal{T}_j): (\boldsymbol{x},\boldsymbol{y}) \in  \mathcal{C}_{n,1-\alpha} \}$

\paragraph{Exact prediction set} An exact prediction set for  $\boldsymbol{Z}_{n+1}=(\boldsymbol{X}_{n+1}, \boldsymbol{Y}_{n+1})$ - which is independent from and identically distributed to $\boldsymbol{Z}_{1},\dots, \boldsymbol{Z}_{n}$ - is the set $\mathcal{C}_{n,1-\alpha}$ based on $\boldsymbol{Z}_{1},\dots, \boldsymbol{Z}_{n}$ such that
\begin{equation}
\mathbb{P}\left(\boldsymbol{Y}_{n+1} \in \mathcal{C}_{n,1-\alpha}\left(\boldsymbol{X}_{n+1}\right)\right) = 1-\alpha
\label{eq::exact_pred_set}
\end{equation}
for any significance level $\alpha \in (0,1)$ and with $\mathcal{C}_{n,1-\alpha}\left(x\right)$ defined as above.

It is important to notice that the left side of Inequality (\ref{eq::valid_pred_set}) and of Equality (\ref{eq::exact_pred_set}) refers to the unconditional coverage reached by the prediction set, i.e. the probability is taken over the i.i.d. draws $\boldsymbol{Z}_1,\dots,\boldsymbol{Z}_{n+1}$.  In view of this, later in the discussion the term \textit{coverage} will be used to indicate the unconditional coverage and the term \textit{empirical coverage} will be used to indicate an estimate of the coverage.

The article is organized as follows: in Section \ref{sec::conf_pred} we introduce the CP framework; in Section \ref{sec::method} we present the method developed; in Section \ref{sec::sim_study} we discuss three simulation studies aimed at investigating different aspects of the method; in Section \ref{sec::bikemi} we apply our method to a real-world application; in Section \ref{sec::conclusion} we provide an overview of the main findings and sketch directions of future research.

\section{The Conformal Prediction Framework}
\label{sec::conf_pred}

Conformal Prediction is an innovative method to build either valid or exact prediction sets under no assumptions other than exchangeable data \citep{vovk2005algorithmic}. Moreover, the CP framework ensures that valid/exact prediction sets are obtained regardless the sample size $n$ (i.e. not only asymptotically), a fact that allows Conformal Prediction to be used in an extremely wide range of different scenarios. 
In this article we consider the Semi-Off-Line Inductive Conformal framework, also known as \textit{Split Conformal} \citep{papadopoulos2002inductive}, which represents a computationally and methodologically convenient alternative to the original  Transductive framework. Split Conformal approach is characterized by two sub-frameworks: Non-Smoothed Split Conformal framework and Smoothed Split Conformal framework \footnote{since the term `Split Conformal' itself is used to indicate `Non-Smoothed Split Conformal', later in the discussion we will use the following two terms to indicate the two sub-frameworks: Split Conformal, Smoothed Split Conformal}. The two procedures are defined below.

\paragraph{Split Conformal method} 
Let $\boldsymbol{z}_1,\dots,\boldsymbol{z}_n$ be realizations of $\boldsymbol{Z}_1,\dots,\boldsymbol{Z}_n$, and let $\{1,\dots, n\}$ be randomly splitted into two sets $\mathcal{I}_1, \mathcal{I}_2$ of size $m$ and $l$ respectively such that $n=m+l$, $m,l \in \mathbb{N}_{>0}$. Let us also define the set $\{ \boldsymbol{z}_h : h \in \mathcal{I}_1\}$ as \textit{training set}, the set $\{ \boldsymbol{z}_d : d \in \mathcal{I}_2\}$ as \textit{calibration set} and the \textit{nonconformity measure}, which represents the key aspect of Conformal Prediction, as any measurable function $A(\{\boldsymbol{z}_h: h \in  \mathcal{I}_1 \} ,\boldsymbol{z})$ taking values in $\bar{\mathbb{R}}$. The Split Conformal approach defines the prediction set for $\boldsymbol{Y}_{n+1}$ as $\mathcal{C}_{n, 1-\alpha}\left(\boldsymbol{x}_{n+1}\right):= \left\{\boldsymbol{y} \in  \prod_{j=1}^p L^{\infty}(\mathcal{T}_j): \delta_{\boldsymbol{y}}>\alpha \right\}$, with
\begin{equation*}
\delta_{\boldsymbol{y}} :=  \frac{\left|\left\{d \in  \mathcal{I}_2 \cup \{n+1\} : R_{d} \geq R_{n+1}\right\}\right|}{l+1},
\end{equation*}
and \textit{nonconformity scores} $R_d:=A( \{\boldsymbol{z}_h: h \in  \mathcal{I}_1 \} ,\boldsymbol{z}_{d})$ for $d \in \mathcal{I}_2$, $R_{n+1}:=A( \{\boldsymbol{z}_h: h \in  \mathcal{I}_1 \} ,\left(\boldsymbol{x}_{n+1},\boldsymbol{y}\right))$. Intuitively, nonconformity score $R_d$ ($R_{n+1}$ respectively) scores how different $\boldsymbol{z}_d$ ($\left(\boldsymbol{x}_{n+1},\boldsymbol{y}\right)$ respectively) is from the training set, and so  $\delta_{\boldsymbol{y}}$ indicates the conformity of $\left(\boldsymbol{x}_{n+1},\boldsymbol{y}\right)$ to the training set compared to the conformity of the elements of the calibration set to the same training set \citep[i.e. it is the p-value of $\left(\boldsymbol{x}_{n+1},\boldsymbol{y}\right)$,][]{vovk2005algorithmic}. 

The Split Conformal method is particulary appealing since it outputs - by construction - finite-sample, valid prediction sets by only assuming exchangeable data. In fact, \citet{diquigiovanni2021importance} show that, under the mild assumption that $\{R_d: d \in  \mathcal{I}_2 \}$ have a continuous joint distribution (an assumption that we will made hereafter), the coverage ensured by Split Conformal prediction set is equal to an easy-to-compute fixed quantity, i.e. $\mathbb{P}\left(\boldsymbol{Y}_{n+1} \in \mathcal{C}_{n,1-\alpha}\left(\boldsymbol{X}_{n+1}\right)\right) =1-\frac{\lfloor (l+1)\alpha \rfloor}{l+1}$, and it is not only greater than or equal to $1-\alpha$. As a consequence, exact (and not only valid) prediction sets are automatically obtained whenever $\lfloor (l+1)\alpha \rfloor = (l+1)\alpha$.

\paragraph{Smoothed Split Conformal method} Moving from the Split Conformal framework, let us consider a single realization of a uniform random variable in $[0,1]$, called $\tau_{n+1}$. The Smoothed Split Conformal approach defines the prediction set for $\boldsymbol{Y}_{n+1}$ as $\mathcal{C}_{n, 1-\alpha,\tau_{n+1}}\left(\boldsymbol{x}_{n+1}\right):= \left\{\boldsymbol{y} \in  \prod_{j=1}^p L^{\infty}(\mathcal{T}_j): \delta_{\boldsymbol{y},\tau_{n+1}}>\alpha \right\}$, with 
\begin{equation*}
\delta_{\boldsymbol{y},\tau_{n+1}} := \frac{\left|\left\{d \in  \mathcal{I}_2: R_{d} > R_{n+1}\right\}\right| + \tau_{n+1} \left|\left\{d \in  \mathcal{I}_2 \cup \{n+1\}: R_{d} = R_{n+1}\right\}\right|}{l+1}.
\end{equation*}
By introducing the element of randomization $\tau_{n+1}$, the Smoothed Split Conformal method outputs finite-sample, exact prediction sets by only assuming exchangeable data \citep{vovk2005algorithmic}.

In order to avoid redundancy, later in the discussion we will mainly focus on Split Conformal method, but the generalization of the main findings of this article to  
the Smoothed Split Conformal method is reported in Appendix A.

\section{Proposed Methodology}
\label{sec::method}
\subsection{Nonconformity Measure}
\label{sec::NCM}
Moving from \citet{diquigiovanni2021importance}, we propose the following nonconformity measure and nonconformity scores:
\begin{equation}
A^s(\{\boldsymbol{z}_h: h \in  \mathcal{I}_1 \}, \boldsymbol{\tilde{z}})= \sup_{j \in \{1,\dots,p\}} \left( \sup_{t \in \mathcal{T}_j} \left| \frac{\tilde{y}_{j}(t)-[\hat{\mu}^j_{\mathcal{I}_1}(\tilde{x}_{j})](t)}{s_{j,\mathcal{I}_1}(t)}\right| \right)
\label{eq::NCM}
\end{equation}
\begin{align}
R_d^s=& \sup_{j \in \{1,\dots,p\}} \left( \sup_{t \in \mathcal{T}_j} \left| \frac{y_{dj}(t)-[\hat{\mu}^j_{\mathcal{I}_1}(x_{dj})](t)}{s_{j,\mathcal{I}_1}(t)}\right| \right), \quad  d \in  \mathcal{I}_2 \label{eq::NCS_cal}\\
R_{n+1}^s=& \sup_{j \in \{1,\dots,p\}} \left( \sup_{t \in \mathcal{T}_j} \left| \frac{y_j(t)-[\hat{\mu}^j_{\mathcal{I}_1}(x_{n+1,j})](t)}{s_{j,\mathcal{I}_1}(t)}\right| \right) \nonumber
\end{align}
with $\boldsymbol{\tilde{z}}=(\boldsymbol{\tilde{x}},\boldsymbol{\tilde{y}})$, $\boldsymbol{\tilde{y}}=(\tilde{y}_1,\dots,\tilde{y}_p)$, $\boldsymbol{\tilde{x}}=(\tilde{x}_1,\dots, \tilde{x}_p)$, $y_j$ the $j$-th component of $\boldsymbol{y}$, $[\hat{\mu}^j_{\mathcal{I}_1}(x_{dj})](t)$ estimate of $[\mu^j(x_{dj})](t)$ based on $\{\boldsymbol{z}_h: h \in  \mathcal{I}_1 \}$, $s_{\mathcal{I}_1}= \{ s_{j,\mathcal{I}_1} \}_{j=1}^p$ \textit{set of modulation functions} with $s_{j,\mathcal{I}_1}: \mathcal{T}_j \rightarrow \mathbb{R}_{>0}$ a (strictly positive) function belonging to $L^{\infty}(\mathcal{T}_j)$ based on  $\{\boldsymbol{z}_h: h \in  \mathcal{I}_1 \}$ called modulation function, and with  the superscript $s$ introduced in order to  emphasize the role of $s_{\mathcal{I}_1}$. It is fundamental to notice that no specific assumptions are made on the estimators $[\hat{\mu}^1_{\mathcal{I}_1}(\cdot)](t),\dots, [\hat{\mu}^p_{\mathcal{I}_1}(\cdot)](t)$ (considered in this case as random variables instead of observed values) since the Conformal framework only requires the nonconformity scores $R_d^s$ and $R_{n+1}^s$ to be computed on the basis of the observations belonging to the training set and  on $\boldsymbol{z}_d$ and $(\boldsymbol{x_{n+1}},\boldsymbol{y})$ respectively. As a consequence, finite-sample, either valid or exact prediction sets are obtained regardless the choice of the regression estimators, allowing Conformal Inference to be satisfactorily performed also when the underlying model is completely misspecified. 

By considering the Split Conformal method and the nonconformity measure (\ref{eq::NCM}), if $\alpha \in (0,1/(l+1) )$ then $\mathcal{C}^s_{n, 1-\alpha}(\boldsymbol{x}_{n+1})= \prod_{j=1}^p L^{\infty}(\mathcal{T}_j)$ since $\delta^s_{\boldsymbol{y}}$ is always greater or equal than $1/(l+1)$. If $\alpha \in [1/(l+1),1 )$ (representing the scenario on which we will focus on hereafter), then 
\begin{align}
\mathcal{C}^s_{n, 1-\alpha}(\boldsymbol{x}_{n+1}):= \bigg\{ \boldsymbol{y} \in \prod_{j=1}^p L^{\infty}(\mathcal{T}_j):  y_j(t) \in \big[ &[\hat{\mu}^j_{\mathcal{I}_1}(x_{n+1,j})](t)-k^s \cdot s_{j,\mathcal{I}_1}(t), \nonumber \\ 
&[\hat{\mu}^j_{\mathcal{I}_1}(x_{n+1,j})](t)+k^s \cdot s_{j,\mathcal{I}_1}(t)] \label{eq:pred_set}\\
& \forall j \in \{1,\dots,p\}, \forall t \in \mathcal{T}_j \bigg\}, \nonumber
\end{align}
with $k^s$ the $\lceil (l+1)(1-\alpha) \rceil$th smallest value in the set $\{ R^s_d: d \in \mathcal{I}_2 \}$. The computation needed to find analytically $\mathcal{C}^s_{n, 1-\alpha}(\boldsymbol{x}_{n+1})$ is provided in Appendix A.1, together with the definition of $\mathcal{C}^s_{n, 1-\alpha,\tau_{n+1}}\left(\boldsymbol{x}_{n+1}\right)$, i.e. the Smoothed Split Conformal prediction set induced by nonconformity measure (\ref{eq::NCM}). 

From a practical point of view, first of all the observed sample $\boldsymbol{z}_1,\dots,\boldsymbol{z}_n$ is used to compute $k^s$ and $s_{1,\mathcal{I}_1},\dots,s_{p,\mathcal{I}_1}$, and after that the prediction set is built around the regression estimates $[\hat{\mu}^j_{\mathcal{I}_1}(x_{n+1,j})](t)$, $j \in \{1,\dots,p\}$. Despite the fact that no specific constraints on $[\hat{\mu}^j_{\mathcal{I}_1}(\cdot)](t)$ are required by the Split Conformal framework, the choice of the regression estimators is fundamental in providing small prediction sets, a key topic that will be investigated in Section \ref{sec::choice_s}: indeed, intuitively one is justified in expecting prediction sets to be smaller when improved regression estimators are chosen since they typically provide smaller nonconformity scores and so a smaller value of $k^s$ \citep{lei2018distribution}. However, later in the discussion (and specifically in Section \ref{sec::sim_study} and Section \ref{sec::bikemi}) we will always consider the regression estimators as given by the application at hand: in fact, our aim is to construct valid/exact prediction sets in general and arbitrary prediction scenarios and not only in specific, well informed frameworks.


Under the exchangeability assumption of the regression pairs and regardless the choice of $s_{\mathcal{I}_1}$ and $[\hat{\mu}^j_{\mathcal{I}_1}(\cdot)](t)$, the prediction sets induced by nonconformity measure (\ref{eq::NCM})
\begin{itemize}
\item are either finite-sample valid (Split Conformal method) or finite-sample exact (Smoothed Split Conformal method) for any distribution $P$;
\item are in closed form;
\item are bands;
\item are scalable \footnote{Indeed, conditional on the computational cost required to calculate the regression estimates and the set of modulation functions (a set that can be chosen to be computationally parsimonious), and by keeping the ratio $l/n$ fixed when $n$ grows, the time required to compute $k^s$ (and therefore to output the prediction set) increases linearly with $l$, and so linearly with $n$}.
\end{itemize}

Note that nonconformity measure (\ref{eq::NCM}) ensures multivariate simultaneous bands, i.e. bands guaranteeing the desired coverage globally (i.e. for the multivariate random function $\boldsymbol{Y}_{n+1}$). 
Proper multivariate simultaneous coverage represents a leap forward with respect to univariate simultaneous coverage (i.e. coverage holding for $Y_{n+1,j}$) 
and pointwise coverage (i.e. coverage holding for $Y_{n+1,j}(t)$). Conformal prediction bands for multivariate functional data (\ref{eq:pred_set}) can be proven to be a superset of the multivariate functional bands found by concatenating  the $p$ univariate prediction bands obtained by applying the nonconformity measure $\sup_{t \in \mathcal{T}_j} \left| \left( y_{j}(t)-[\hat{\mu}^j_{\mathcal{I}_1}(x_{j})](t) \right) / s_{j,\mathcal{I}_1}(t)\right| $ to the $p$ components separately \citep{diquigiovanni2021importance}, and also a superset of the multivariate functional bands found by concatenating the pointwise prediction intervals obtained by applying the pointwise nonconformity measure $ \left| \left( y_{j}(t)-[\hat{\mu}^j_{\mathcal{I}_1}(x_{j})](t) \right) / s_{j,\mathcal{I}_1}(t)\right|$ $\forall j \in \{1,\dots,p\}$, $\forall t \in \mathcal{T}_j$ (see Appendix A.1 for the proof). In other words, multivariate functional simultaneous bands (\ref{eq:pred_set})  ensure also both univariate simultaneous and pointwise validity, while the converse is not guaranteed. The topic is further addressed by means of a simulation study in Section \ref{sec::sim_study2}.

Alongside the choice to base the nonconformity measure on the supremum metric, the set $s_{\mathcal{I}_1}$ of (strictly positive) modulation functions $s_{j, \mathcal{I}_1}$ represents the core of our approach. First of all, one can notice that prediction bands induced by $\{ s_{j,\mathcal{I}_1} \}_{j=1}^p$ and by $\{ \lambda \cdot s_{j,\mathcal{I}_1} \}_{j=1}^p$ coincide $\forall \lambda \in \mathbb{R}_{>0}$ (see Appendix A.1 for the proof), and so later in the discussion we will consider, for any equivalence class, the set of modulation functions such that $\sum_{j=1}^p  \int_{\mathcal{T}_j} s_{j,\mathcal{I}_1}(t)  dt=1$. In the next Section, we detail the role of $s_{\mathcal{I}_1}$ by highlighting its impact on the efficiency (i.e. the size) of the prediction bands and we propose a specific set of modulation functions able to guarantee an asymptotic result in terms of efficiency. 

\subsection{The Choice of the Set of Modulation Functions}
\label{sec::choice_s}

\begin{figure}
\begin{center}
\includegraphics[width=11.7cm,height=9.045cm]{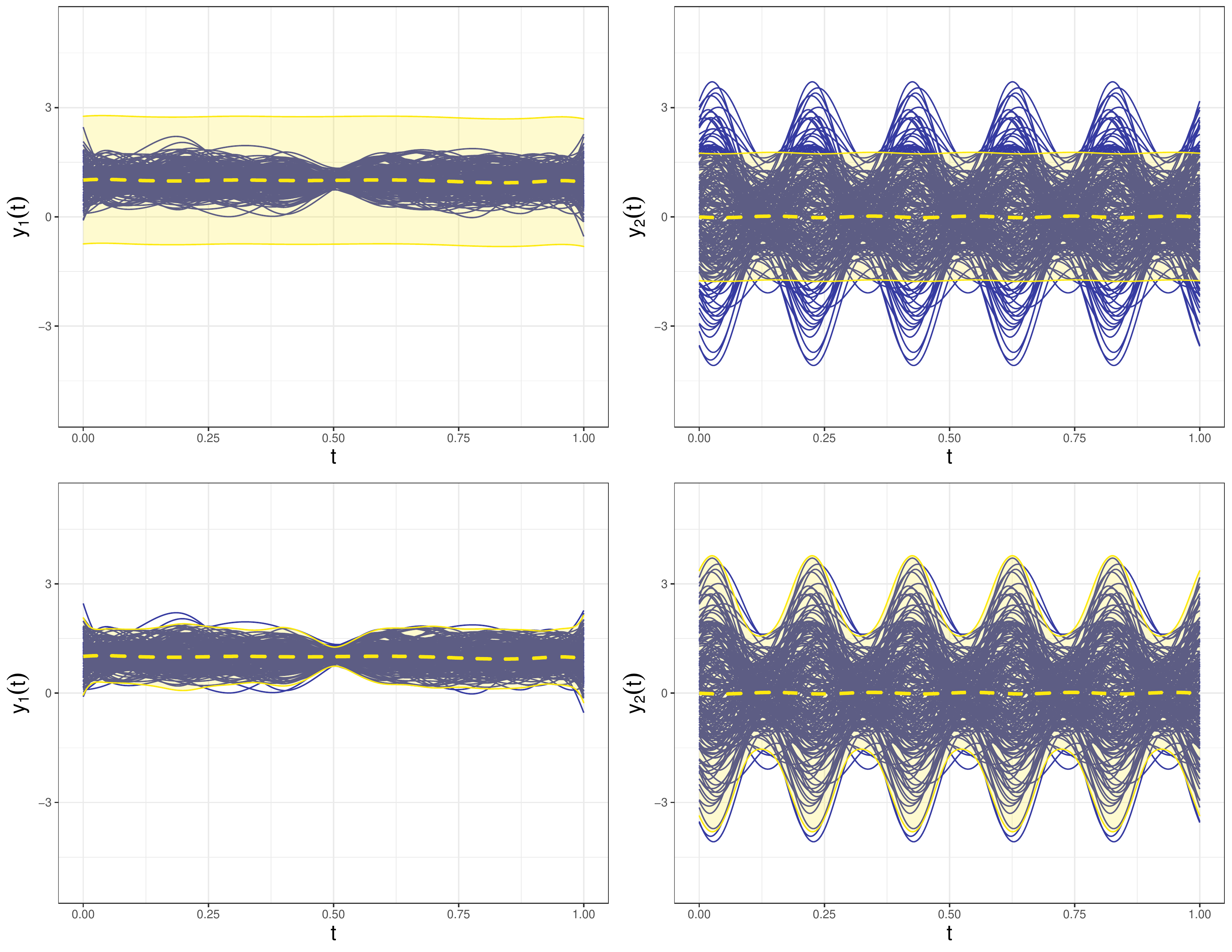} 
\end{center}
\caption{Split Conformal multivariate prediction band  for $\boldsymbol{Y}_{201}=(Y_{201,1},Y_{201,2})$  obtained by considering $\{ s^0_{j}\}_{j=1}^2$ (at the top) and $\{ s^{\sigma}_{j,\mathcal{I}_1}\}_{j=1}^2$ (at the bottom) as set of modulation functions. The dashed yellow lines represent the regression estimates. $\alpha=0.25$, $n=200$, $m=l=100$.  \label{fig::stdev_vs_id_mult} }
\end{figure}

Intuitively, in addition to the appealing properties presented in Section \ref{sec::NCM}, a prediction band should modulate its width over $\mathcal{T}_1,\dots,\mathcal{T}_p$ according to the local variability of the data. Specifically, the aim is to obtain prediction bands able to properly manage the fact that: focusing on the $j$-th component, the pointwise evaluations of functional data may be characterized by highly different variability when $t \in \mathcal{T}_j$ varies; the $p$ components may be characterized by different magnitude. In order to achieve these two purposes, a careful choice of a data-driven set of modulation functions $s_{\mathcal{I}_1}$ is recommended. In order to clarify this concept, let us consider the following example: let $p=2$ with $\boldsymbol{y}_1,\dots, \boldsymbol{y}_{200}$ independent realizations of $\boldsymbol{Y}_1,\dots, \boldsymbol{Y}_{200}$ such that $Y_{i1}(t)=\beta_1(t) + \varepsilon_{i1}(t)$ and $Y_{i2}(t)=\beta_2(t) + \varepsilon_{i2}(t)$ ($i=1,\dots,200,$ $\mathcal{T}_1 = \mathcal{T}_2 = [0,1]$), with the systematic components defined simply as $\beta_1(t) =1$,  $\beta_2(t) =0$ $\forall t \in [0,1]$ and the independent functional error components $\{ \varepsilon_{i1} \}_{i=1}^{200}$ ($\{ \varepsilon_{i2} \}_{i=1}^{200}$ respectively) obtained by means of a B-spline basis expansion (Fourier basis expansion respectively) with normally distributed random vectors as coefficients. In full generality, we consider $[\hat{\mu}^j_{\mathcal{I}_1}( x_{n+1,j})](t)=\hat{\beta}_j(t)$, $j=1,2$, with $\hat{\beta}_1(t),\hat{\beta}_2(t)$ the estimates (based on $ \{\boldsymbol{z}_h: h \in  \mathcal{I}_1 \}$) obtained by fitting the two concurrent functional-on-functional linear models \citep{ramsay_functional_2005}. This example represents the simplest, almost trivial regression scenario which allows to - hopefully - easily understand the crucial role of $s_{\mathcal{I}_1}$, but the discussion presented hereafter naturally holds also when decidedly more complex regression functions and regression estimators are taken into account. Figure \ref{fig::stdev_vs_id_mult} 
shows the multivariate prediction band for $\boldsymbol{Y}_{201}=(Y_{201,1},Y_{201,2})$ obtained by considering two different sets of modulation functions: the two panels at the top of the Figure \ref{fig::stdev_vs_id_mult} show the multivariate prediction band obtained by not modulating (i.e. by setting $s_{1,\mathcal{I}_1}(t)=s_{2,\mathcal{I}_1}(t)=1/\sum_{j=1}^2  \left| \mathcal{T}_j \right| \propto 1$ $\forall t \in [0,1]$), whereas the two panels at the bottom of  the same Figure show the prediction band obtained by considering the two standard deviation functions of the functional residuals as modulation functions  (after normalization in order to meet the condition $\sum_{j=1}^2  \int_{\mathcal{T}_j} s_{j,\mathcal{I}_1}(t)  dt=1$). In order to distinguish the two sets of modulation functions, later in the discussion we will denote the first set by $ s^0:=\{ s^0_{j}\}_{j=1}^p$ (whose notation excludes the subscript $\mathcal{I}_1$ to remark its lack of dependence on the training set) and the second one by $s^{\sigma}_{\mathcal{I}_1}:=\{ s^{\sigma}_{j,\mathcal{I}_1}\}_{j=1}^p$. The prediction sets are obtained by considering the Split Conformal framework and by setting $\alpha=0.25$, $m=l=100$. Focusing on the two panels at the top of Figure \ref{fig::stdev_vs_id_mult}, it is possible to notice that the two univariate prediction bands
are far from desirable: specifically, the univariate prediction band related to $Y_{201,1}$ is large along all the domain $\mathcal{T}_1$, whereas the one related to $Y_{201,2}$ contains almost all the pointwise evaluations of the functional data in the low-variance parts of $\mathcal{T}_2$ but excludes many pointwise evaluations in the other, high-variance parts of the domain. In this specific case, the absence of a modulation process does not allow to take into account: first of all, the different variability of the data over $\mathcal{T}_1$ and $\mathcal{T}_2$ respectively; secondly, the different magnitude that characterizes the two components. In so doing, one is justified in expecting that a procedure based on $\{ s^0_{j}\}_{j=1}^2$, although able to output a valid prediction band, may be of limited practical use in real applications. Vice versa, the set of modulation functions $\{ s^{\sigma}_{j,\mathcal{I}_1}\}_{j=1}^2$ properly adapts the width of the prediction band according to the local variability of functional data, allowing for a meaningful, interpretable and useful prediction band.  

Beyond these common-sense considerations, a criterion that is both reasonable and well-established in Conformal Prediction to discriminate between procedures able to guarantee validity is the minimization of the size of the prediction sets outputted \citep[also known, in the Conformal framework, as maximization of efficiency,][]{balasubramanian2014conformal}: this choice is due to the fact that desirable prediction sets should include subsets of the sample space where the probability mass is highly concentrated \citep{lei2013distribution}. 
In the context of our article, the aim would be to find the nonconformity measure $A^s(\{\boldsymbol{z}_h: h \in  \mathcal{I}_1 \}, \cdot)$ (and so, practically, the set of modulation functions $s_{\mathcal{I}_1}$) inducing the smallest prediction bands. The first, fundamental step in assessing the size of a prediction band for multivariate functional data is the definition of the concept of `size', a nontrivial task if compared to the traditional univariate and multivariate statistical settings. By generalizing the definition given in \citet{diquigiovanni2021importance} to the multivariate case, we define the size of a multivariate prediction band as the sum of the $p$ areas between the upper and lower bound of the $p$ univariate prediction bands: 
\begin{equation}
\mathcal{Q}(s_{\mathcal{I}_1}):=\sum_{j=1}^p \int_{\mathcal{T}_j} 2 \cdot k^s \cdot s_{j, \mathcal{I}_1}(t) dt =2 \cdot k^s.
\label{eq:area_multiv}
\end{equation}

Since $\mathcal{Q}(s_{\mathcal{I}_1})$ is a random variable depending on $\boldsymbol{Z}_1,\dots,\boldsymbol{Z}_n$, the task of finding the set of modulation functions minimizing the risk functional $\mathbb{E}[\mathcal{Q}(s_{\mathcal{I}_1})]$ is unfeasible in the case of no assumptions on $P$. A simplification of such a complex task consists of considering the quantity to be minimized $k^s (\propto \mathcal{Q}(s_{\mathcal{I}_1}))$ as an observed value depending on $\boldsymbol{z}_1,\dots,\boldsymbol{z}_n$ instead of on $\boldsymbol{Z}_1,\dots,\boldsymbol{Z}_n$ according to the empirical risk minimization principle \citep{vapnik1992principles}. In so doing, the optimization problem is certainly simplified, but its resolution still remains unfeasible due to the specific structure of $k^s$. Indeed, $k^s$ is a specific empirical quantile of $\{ R^s_d: d \in \mathcal{I}_2 \}$, and $R^s_d$ (see Equation (\ref{eq::NCS_cal})) depends by construction both on the training set through $\{ \boldsymbol{z}_h : h \in \mathcal{I}_1\}$ and on the calibration set through $\boldsymbol{z}_d$. Since by construction the set of modulation functions $s_{\mathcal{I}_1}$ depends only on the training set (as its dependence on the calibration set would imply not to obtain closed-form valid prediction bands), no rule minimizing $k^s$ only by combining the elements of the training set (i.e. by varying $s_{\mathcal{I}_1}$) can be found for general $\boldsymbol{z}_1,\dots,\boldsymbol{z}_n$.

In view of this, we propose an alternative, unconventional strategy to build a set of modulation functions able to guarantee an asymptotic result in terms of efficiency. Specifically, the purpose is to find a couple of sets of functions $(\bar{s}_{\mathcal{I}_1}, \bar{s}^{c}_{\mathcal{I}_1,\mathcal{I}_2})$ such that:
\begin{itemize}
\item $\bar{s}^{c}_{\mathcal{I}_1,\mathcal{I}_2}:=\{\bar{s}^{c}_{j,\mathcal{I}_1,\mathcal{I}_2}\}_{j=1}^p$ is a set of functions such that $\bar{s}^{c}_{j,\mathcal{I}_1,\mathcal{I}_2}$ meets the definition of modulation function, but depends also on the calibration set through $\{ \boldsymbol{z}_d : d \in \mathcal{I}_2\}$, $\forall j \in \{1,\dots,p\}$
\item  prediction bands obtained by using $\bar{s}^{c}_{\mathcal{I}_1,\mathcal{I}_2}$ as set of modulation functions are smaller than or equal to (in terms of Equation (\ref{eq:area_multiv})) those induced by the set of modulation functions $s^0$ for every possible value of $n$ and for every possible observed sample $\boldsymbol{z}_1,\dots,\boldsymbol{z}_n$
\item $\bar{s}_{\mathcal{I}_1}=\{\bar{s}_{j,\mathcal{I}_1} \}_{j=1}^p$ is a set of modulation functions such that $\bar{s}_{j,\mathcal{I}_1}$ is equal to $\bar{s}^{c}_{j,\mathcal{I}_1,\mathcal{I}_2}$, but in which the dependence on $\{ \boldsymbol{z}_d : d \in \mathcal{I}_2\}$ is replaced by the dependence on $\{ \boldsymbol{z}_h : h \in \mathcal{I}_1\}$ $\forall j \in \{1,\dots,p\}$
\item $\bar{s}^{c}_{j,\mathcal{I}_1,\mathcal{I}_2}$ and $\bar{s}_{j,\mathcal{I}_1}$ converge to the same function when $m,l \to +\infty$, $\forall j \in \{1,\dots,p\}$
\end{itemize}

In so doing, prediction bands induced by the set of modulation functions $\bar{s}_{\mathcal{I}_1}$ are characterized by all the appealing properties presented in Section \ref{sec::NCM} (including validity) and are asymptotically not wider than those induced by  $s^0$ regardless the specific sample $\boldsymbol{z}_1,\dots,\boldsymbol{z}_n$. In order to find $(\bar{s}_{\mathcal{I}_1}, \bar{s}^{c}_{\mathcal{I}_1,\mathcal{I}_2})$ satisfying the aforementioned conditions, let us consider the structure of $k^s$: operationally, $k^s$ computes a summary of the multivariate functional residual 
for every observation in the calibration set, and selects the $\lceil (l+1)(1-\alpha) \rceil$th smallest value among them. In particular: the summary is naturally induced by the specific nonconformity measure used, which searches the greatest value of the absolute value of the modulated multivariate functional residual over the $p$ domains $\mathcal{T}_1,\dots,\mathcal{T}_p$; $k^s$ is not affected by the $l-\lceil (l+1)(1-\alpha) \rceil$ greatest values of $\{ R^s_d: d \in \mathcal{I}_2 \}$. In view of this, a proper candidate for $\bar{s}^{c}_{\mathcal{I}_1,\mathcal{I}_2}$ should ignore the elements of $\{ \boldsymbol{z}_d : d \in \mathcal{I}_2\}$ leading to the $l-\lceil (l+1)(1-\alpha) \rceil$ greatest values of $\{ R^s_d: d \in \mathcal{I}_2 \}$ and should modulate data based on the most extreme value observed $\forall t \in \mathcal{T}_j, j \in \{1,\dots,p\}$.

Therefore, the couple of sets of functions $(\bar{s}_{\mathcal{I}_1}, \bar{s}^{c}_{\mathcal{I}_1,\mathcal{I}_2})$ we propose - which represents a generalization of the finding of \citet{diquigiovanni2021importance} in the univariate case - is defined below. Formally, the set of functions $\bar{s}^{c}_{\mathcal{I}_1,\mathcal{I}_2}$ is such that
\begin{equation*}
 \bar{s}^{c}_{j,\mathcal{I}_1,\mathcal{I}_2}(t):=\frac{\max_{d \in \mathcal{H}_2} |y_{dj}(t)-[\hat{\mu}^j_{\mathcal{I}_1}(x_{dj})](t) |}{\sum_{j=1}^p \int_{\mathcal{T}_j}\max_{d \in \mathcal{H}_2} |y_{dj}(t)-[\hat{\mu}^j_{\mathcal{I}_1}(x_{dj})](t) | dt}
\end{equation*}
$\forall$ $j=1,\dots,p, t \in \mathcal{T}_j$ with 
\begin{equation*}
\mathcal{H}_2:=\Big\{d \in \mathcal{I}_2: \sup_{j \in \left\{1,\dots,p\right\}} \left(\sup_{t \in \mathcal{T}_j} |y_{dj}(t)-[\hat{\mu}^j_{\mathcal{I}_1}(x_{dj})](t) |\right) \leq k \Big\}
\end{equation*}
and $k=k^{s^0}/\sum_{j=1}^p  \left| \mathcal{T}_j \right|$  the $\lceil (l+1)(1-\alpha) \rceil$th smallest value in the set $$\Big\{\sup_{j \in \left\{1,\dots,p\right\}} \left( \sup_{t \in \mathcal{T}_j} \left| y_{dj}(t)-[\hat{\mu}^j_{\mathcal{I}_1}(x_{dj})](t) \right| \right) : d \in \mathcal{I}_2 \Big\}.$$ 
For the sake of simplicity, we assumed $\max_{d \in \mathcal{H}_2} |y_{dj}(t)-[\hat{\mu}^j_{\mathcal{I}_1}(x_{dj})](t) | \neq 0$ $\forall j \in \{1,\dots,p\}, t \in \mathcal{T}_j$. If this condition does not hold for at least one couple $(t,j)$ but the condition $\sum_{j=1}^p \int_{\mathcal{T}_j}\max_{d \in \mathcal{H}_2} |y_{dj}(t)-[\hat{\mu}^j_{\mathcal{I}_1}(x_{dj})](t) | dt \neq 0$ still holds (the case in which $\sum_{j=1}^p \int_{\mathcal{T}_j}\max_{d \in \mathcal{H}_2} |y_{dj}(t)-[\hat{\mu}^j_{\mathcal{I}_1}(x_{dj})](t) | dt = 0$ represents a pathological case of no practical interest), in order to have that $\bar{s}^{c}_{j,\mathcal{I}_1,\mathcal{I}_2}(t)>0$ $\forall j=1,\dots,p, t \in \mathcal{T}_j$ it is sufficient to add a small, positive value to $\bar{s}^{c}_{j,\mathcal{I}_1,\mathcal{I}_2}(t)$ and to normalize accordingly.

The set of modulation functions $\bar{s}_{\mathcal{I}_1}$ is such that
\begin{equation*}
 \bar{s}_{j, \mathcal{I}_1}(t):=\frac{\max_{h \in \mathcal{H}_1}|y_{hj}(t)-[\hat{\mu}^j_{\mathcal{I}_1}(x_{hj})](t) |}{\sum_{j=1}^p \int_{\mathcal{T}_j}\max_{h \in \mathcal{H}_1} |y_{hj}(t)-[\hat{\mu}^j_{\mathcal{I}_1}(x_{hj})](t) | dt}
\end{equation*}
$\forall$ $j=1,\dots,p, t \in \mathcal{T}_j$ with $\mathcal{H}_1=\mathcal{I}_1$ if $\lceil (m+1)(1-\alpha) \rceil > m$, otherwise
\begin{equation*}
\mathcal{H}_1:=\Big\{h \in \mathcal{I}_1: \sup_{j \in \left\{1,\dots,p\right\}} \left( \sup_{t \in \mathcal{T}_j} |y_{hj}(t)-[\hat{\mu}^j_{\mathcal{I}_1}(x_{hj})](t) | \right) \leq \gamma \Big\} 
\end{equation*}
and $\gamma$ the $\lceil (m+1)(1-\alpha) \rceil$th smallest value in the set $$\Big\{ \sup_{j \in \left\{1,\dots,p\right\}} \left( \sup_{t \in \mathcal{T}_j} \left| y_{hj}(t)-[\hat{\mu}^j_{\mathcal{I}_1}(x_{hj})](t) \right| \right) : h \in \mathcal{I}_1 \Big\}.$$

If $\exists (t,j)$ such that $\max_{h \in \mathcal{H}_1}|y_{hj}(t)-[\hat{\mu}^j_{\mathcal{I}_1}(x_{hj})](t) |=0$, the adjustment used for $ \bar{s}^{c}_{j,\mathcal{I}_1,\mathcal{I}_2}$ is implemented. 

Specifically, the fact that the set of modulation functions $\bar{s}_{\mathcal{I}_1}$ depends on $\alpha$ (through $\gamma$) allows for a procedure able to modulate data according to the specific value $1-\alpha$, i.e. the desired nominal coverage. In addition, such an unconventional set of modulation functions is particularly useful when functional residuals show a non-standard behavior (e.g. there are outliers).  The following two theorems show that $(\bar{s}_{\mathcal{I}_1}, \bar{s}^{c}_{\mathcal{I}_1,\mathcal{I}_2})$ satisfies the aforementioned conditions.

\begin{theorem}
\label{th:shared_convergence}
 Let $m/n = \theta$ with $0 < \theta < 1$ and let $\mathrm{Var}\big[[\hat{\mu}^j_{\mathcal{I}_1}(X_{ij})](t)\big] \to 0$ $\forall i \in\{1,\dots,n\}$, $\forall t \in \mathcal{T}_j$, $\forall j \in \{1,\dots,p\}$  when $m \to +\infty$. Then  $\bar{s}^{c}_{j,\mathcal{I}_1,\mathcal{I}_2}$ and $ \bar{s}_{j,\mathcal{I}_1}$ converge to the same function $\forall j \in \{1,\dots,p\}$ when $n \to +\infty$.
\end{theorem}

\begin{theorem}
\label{th:better_than_not_modul}
If at least one of the functions $\{ \bar{s}^c_{j, \mathcal{I}_1,\mathcal{I}_2}(t)\}_{j=1}^p$ is not constant almost everywhere over its domain, then $\mathcal{Q}(s^{0}) > \mathcal{Q}( \bar{s}^{c}_{\mathcal{I}_1,\mathcal{I}_2})$. Otherwise, $\mathcal{Q}(s^{0}) = \mathcal{Q}( \bar{s}^{c}_{\mathcal{I}_1,\mathcal{I}_2})$.
\end{theorem}

See Appendix A.2 for both proofs, together with the generalization of $(\bar{s}_{\mathcal{I}_1}, \bar{s}^{c}_{\mathcal{I}_1,\mathcal{I}_2})$, Theorem \ref{th:shared_convergence} and Theorem \ref{th:better_than_not_modul} to the Smoothed Split Conformal framework. Due to the very mild conditions required by the two theorems to hold, the set of modulation functions $ \bar{s}_{\mathcal{I}_1}$ can be used in many general frameworks and provides a new, we believe appealing data-driven alternative to other solutions (e.g. $s^{\sigma}_{\mathcal{I}_1})$. In the next Section, the set of modulation functions $\bar{s}_{\mathcal{I}_1}$ is compared to other sets of modulation functions in different simulated scenarios. 

\section{Simulation Study}
\label{sec::sim_study}

In this Section we perform three simulation studies aimed at evaluating different practical aspects of the method presented in Section \ref{sec::method}. 
Since, to our knowledge, there are no methods dealing with building prediction bands in a multivariate functional setting, the simulations will focus on exploring the empirical properties of our method. In Section \ref{sec::sim_study1}, the empirical coverage provided by the prediction bands is evaluated in different scenarios, considering different sample sizes and different kinds of model misspecification. In Section \ref{sec::sim_study2}, we compare the multivariate prediction bands obtained by the method presented in this article with those obtained by concatenating the $p$ univariate prediction bands induced by the Conformal approach of \citet{diquigiovanni2021importance}. Finally,  in Section \ref{sec::sim_study3}, the three sets of modulation functions presented in Section \ref{sec::method} $(\{ s^{0}_{j}\}_{j=1}^p, \{ s^{\sigma}_{j,\mathcal{I}_1}\}_{j=1}^p, \{ \bar{s}_{j,\mathcal{I}_1}\}_{j=1}^p)$ are compared in terms of efficiency in order to highlight their strengths and weaknesses.

In all simulation studies, some quantities are kept fixed: $p=2$, $\mathcal{T}_1=\mathcal{T}_2=[0,1]$, $\alpha=0.10$. Three possible sample sizes are taken into account: $n=20, n=200, n=2000$.   We focus on the Split Conformal method and since the coverage reached by Split Conformal prediction set is $1-\lfloor (l+1)\alpha \rfloor/(l+1)$ (see Section \ref{sec::conf_pred}), the size of the calibration set is set equal to $l=9,l=99,l=999$ respectively in order to obtain $1-\lfloor (l+1)\alpha \rfloor/(l+1)=1-\alpha$ and consequently to facilitate the readability of the results. A possible alternative would be to consider a different value of $l$ (e.g. $n/2$) and to evaluate the empirical coverage taking into account the coverage $1-\lfloor (l+1)\alpha \rfloor/(l+1)$. Each combination of simulation study, scenario, sample size, regression estimators and set of modulation functions is evaluated based on $N=5000$ replications. Specifically, for each replication, a sample $\boldsymbol{z}_1,\dots,\boldsymbol{z}_{n+1}$ is generated and $n$ randomly chosen elements are assigned to the training and calibration sets, whereas the remaining element is considered as the one we aim to predict (however, for the sake of simplicity, hereafter we will simply define the two sets as $\{ \boldsymbol{z}_i\}_{i=1}^n$ and $\boldsymbol{z}_{n+1}$). All simulations are computed using the R Programming Language \citep{Rcoreteam}.

\subsection{Simulation Study 1: Coverage}
\label{sec::sim_study1}

The aim of the simulation study in this Section is to evaluate the empirical coverage (computed as the fraction of the $N=5000$ replications in which $\boldsymbol{y}_{n+1}$ belongs to  $\mathcal{C}_{n,1-\alpha}\left(\boldsymbol{x}_{n+1}\right)$) reached by the method presented in Section \ref{sec::method} in different scenarios and for different values of $n$.

Specifically, the simulation study consists of two scenarios. In the first one, the systematic component generating data is linear and, in addition to the case in which the model is correctly specified, two different kinds of model misspecification are taken into account: misspecification due to omitted relevant variable  and misspecification due to inclusion of irrelevant variable \citep[see][]{rao1971some}. In the second scenario, a third kind of model misspecification is evaluated, i.e. functional form misspecification \citep[see][]{wooldridge1994simple}. The two scenarios are formally defined as follows:
\begin{itemize}
\item \textit{Scenario 1} 

\begin{align*}
Y_{i1}(t)=&\beta_0(t) + \beta_1(t) w_{i}+ \varepsilon_{i1}(t), \quad i \in \{1,\dots,n+1\}, t \in [0,1] \\
Y_{i2}(t)=&\beta_0(t) + \beta_2(t) w_{i}^2+ \varepsilon_{i2}(t), \quad i \in \{1,\dots,n+1\}, t \in [0,1] 
\end{align*}
with $w_i=i/(n+1)$, $\beta_0(t), \beta_1(t), \beta_2(t)$ generated by means of a B-spline basis expansion of order four, with six basis functions, equally spaced knots, coefficients generated independently by a standard normal random variable and $\varepsilon_{i1}(t), \varepsilon_{i2}(t)$ independent functional errors obtained by means of the same B-spline basis expansion with independent standard normal random variables as coefficients. It is important to notice that regression coefficient functions $\beta_0,\beta_1,\beta_2$ are generated only once, i.e. they do not vary between the $N=5000$ replications.  
\item \textit{Scenario 2}

\begin{align*}
Y_{i1}(t)=&\exp(\beta_0(t) + \beta_1(t) w_{i}+ \varepsilon_{i1}(t)), \quad i \in \{1,\dots,n+1\}, t \in [0,1] \\
Y_{i2}(t)=&\exp(\beta_0(t) + \beta_2(t) w_{i}^2+ \varepsilon_{i2}(t)), \quad i \in \{1,\dots,n+1\}, t \in [0,1] 
\end{align*}
with $w_i, \beta_0(t), \beta_1(t), \beta_2(t),\varepsilon_{i1}(t), \varepsilon_{i2}(t)$ defined as in Scenario 1.
\end{itemize}
Both scenarios are evaluated considering the following three regression estimates: 
\begin{itemize}
\item \textit{Set of Covariates 1}. $[\hat{\mu}^1_{\mathcal{I}_1}( x_{i,1}=\{1\})](t)=[\hat{\mu}^2_{\mathcal{I}_1}( x_{i,2}=\{1\})](t)=\hat{\beta}_0(t)$ 
\item \textit{Set of Covariates 2}. $[\hat{\mu}^1_{\mathcal{I}_1}( x_{i,1}=\{1,w_i\})](t)=\hat{\beta}_0(t) + \hat{\beta}_1(t) w_i$ and $[\hat{\mu}^2_{\mathcal{I}_1}( x_{i,2}=\{1,w^2_i\})](t)=\hat{\beta}_0(t) + \hat{\beta}_2(t) w^2_i$ 
\item \textit{Set of Covariates 3}. $[\hat{\mu}^1_{\mathcal{I}_1}( x_{i,1}=\{1,w_i,w_i^2\})](t)=[\hat{\mu}^2_{\mathcal{I}_1}( x_{i,2}=\{1,w_i,w_i^2\})](t)=\hat{\beta}_0(t) + \hat{\beta}_1(t) w_i + \hat{\beta}_2(t) w^2_i$ 
\end{itemize}
with $\hat{\beta}_0(t), \hat{\beta}_1(t), \hat{\beta}_2(t)$ the estimates (based on $ \{\boldsymbol{z}_h: h \in  \mathcal{I}_1 \}$) obtained by fitting each time the corresponding functional-on-scalar linear model. Focusing on Scenario 1, `Set of Covariates 1' represents the  omitted relevant variable case,  `Set of Covariates 2' represents the  case in which the model is correctly specified and `Set of Covariates 3' represents the case in which an irrelevant variable is included, whereas Scenario 2 is characterized by functional form misspecification. 

Table \ref{tab:validity_simstudy1} 
\begin{table}[t]
\begin{center}
\centering
\begin{tabular}{|ccccc c c c c|}
\hline
\multicolumn{1}{|c}{}&\multicolumn{3}{c|}{\cellcolor{light-gray}\textbf{Scenario 1}}\\
\multicolumn{1}{|c}{}&\cellcolor{light-light-gray}Set of Cov. 1&\cellcolor{light-light-gray}Set of Cov. 2&\multicolumn{1}{c|}{\cellcolor{light-light-gray}Set of Cov. 3}\\ 
$\cellcolor{light-light-gray} n=20$&0.894[0.886,0.903]&0.896[0.888,0.905] &\multicolumn{1}{c|}{0.904[0.896,0.912]} \\  
$\cellcolor{light-light-gray}n=200$&0.902[0.894,0.910]&0.894[0.885,0.903] &\multicolumn{1}{c|}{0.901[0.893,0.909]} \\  
$\cellcolor{light-light-gray}n=2000$&0.899[0.890,0.907]&0.906[0.898,0.914] &\multicolumn{1}{c|}{0.902[0.894,0.911]} \\  \hline
\multicolumn{1}{|c}{}&\multicolumn{3}{c|}{\cellcolor{light-gray}\textbf{Scenario 2}}\\
\multicolumn{1}{|c}{}&\cellcolor{light-light-gray}Set of Cov. 1&\cellcolor{light-light-gray}Set of Cov. 2&\multicolumn{1}{c|}{\cellcolor{light-light-gray}Set of Cov. 3}\\ 
$\cellcolor{light-light-gray}n=20$&0.907[0.899,0.915]&0.899[0.890,0.907] &\multicolumn{1}{c|}{0.904[0.896,0.913]} \\  
$\cellcolor{light-light-gray}n=200$&0.899[0.891,0.907]&0.898[0.890,0.907] &\multicolumn{1}{c|}{0.901[0.893,0.909]} \\  
$\cellcolor{light-light-gray}n=2000$&0.893[0.884,0.901]&0.893[0.884,0.901]&\multicolumn{1}{c|}{0.899[0.891,0.908]} \\  \hline
\end{tabular}

\caption{Simulation study 1: empirical coverage and related 95\% confidence interval in brackets for each combination of scenario, sample size and set of covariates. $\alpha=0.10$, set of modulation functions $\{ s^{\sigma}_{j,\mathcal{I}_1}\}_{j=1}^2$.}
\label{tab:validity_simstudy1}

\end{center}
\end{table}
shows the empirical coverage $\hat{p}$, as well as the 95\% confidence interval $[\hat{p} \pm 1.96 \sqrt{\hat{p}(1-\hat{p})/N} ]$, obtained for each combination of scenario, sample size and set of covariates considering the set of modulation functions $\{ s^{\sigma}_{j,\mathcal{I}_1}\}_{j=1}^2$. The results are decidedly satisfactory, as the empirical coverages are really close to $1-\alpha=0.90$ and the observed confidence intervals always include the desired coverage regardless the specific combination of scenario, sample size and set of covariates considered. Specifically, the method presented in Section \ref{sec::method} is able to guarantee the desired coverage also when the sample size is small and the model misspecified.


\subsection{Simulation Study 2: Univariate and Multivariate Prediction Bands}
\label{sec::sim_study2}

The simulation study of this Section is aimed at comparing 
the Multivariate Prediction Bands outputted by the method presented in this article (\textit{MPB method}) to the bands obtained by Concatenating the $p$ Univariate prediction Bands provided by the Conformal approach presented in \citet{diquigiovanni2021importance} (\textit{CUB method}, see Section \ref{sec::NCM} for further details). We focus on two aspects: empirical coverage and efficiency. The empirical coverage is evaluated as described in Section \ref{sec::sim_study1}, whereas for each of the $N=5000$ replications the size of the observed prediction band $\mathcal{C}_{n,1-\alpha}\left(\boldsymbol{x}_{n+1}\right)$ is defined as the average value $\mathcal{Q}(\cdot)/2$ (see Equation (\ref{eq:area_multiv})).

Two different scenarios are considered: in the first one, the two components $Y_{i,1},Y_{i,2}$ share the systematic component, but they are characterized by independent error terms; in the second one, the two components both share the systematic component and the error term in the first half of the domain. In so doing, one is justified in expecting the method presented in this article not to be affected by the different specification of the error terms in terms of coverage, while the CUB method to provide different empirical coverages according to the scenario considered.  The two scenarios are:
\begin{itemize}
\item \textit{Scenario 1} 

\begin{align*}
Y_{i1}(t)=&\beta_0(t) + \beta_1(t) w_{i} + \beta_2(t) w_{i}^2 + \varepsilon_{i1}(t), \quad i \in \{1,\dots,n+1\}, t \in [0,1] \\
Y_{i2}(t)=&\beta_0(t) + \beta_1(t) w_{i} + \beta_2(t) w_{i}^2+ \varepsilon_{i2}(t), \quad i \in \{1,\dots,n+1\}, t \in [0,1] 
\end{align*}
with $w_i, \beta_0(t), \beta_1(t), \beta_2(t),\varepsilon_{i1}(t), \varepsilon_{i2}(t)$ defined as in Section \ref{sec::sim_study1}.
\item  \textit{Scenario 2} 

\begin{align*}
Y_{i1}(t)=&\beta_0(t) + \beta_1(t) w_{i} + \beta_2(t) w_{i}^2 + \eta_{i1}(t), \quad i \in \{1,\dots,n+1\}, t \in [0,1] \\
Y_{i2}(t)=&\beta_0(t) + \beta_1(t) w_{i} + \beta_2(t) w_{i}^2+ \eta_{i2}(t), \quad i \in \{1,\dots,n+1\}, t \in [0,1] 
\end{align*}
with $\eta_{i1}(t)=\varepsilon_{i1}(t)$, 
\begin{equation*}
\eta_{i2}(t)=\begin{cases}
\varepsilon_{i1}(t) & \text{ $t \in [0,0.5]$}\\ 
\varepsilon_{i2}(t) & \text{$t \in (0.5,1]$}\\ 
\end{cases}
\end{equation*}
 and $w_i, \beta_0(t), \beta_1(t), \beta_2(t),\varepsilon_{i1}(t), \varepsilon_{i2}(t)$ defined as in Section \ref{sec::sim_study1}.
\end{itemize}
As in the previous simulation study, three regression estimates are considered:  
\begin{itemize}
\item \textit{Set of Covariates 1}. $[\hat{\mu}^1_{\mathcal{I}_1}( x_{i,1}=\{1\})](t)=[\hat{\mu}^2_{\mathcal{I}_1}( x_{i,2}=\{1\})](t)=\hat{\beta}_0(t)$ 
\item \textit{Set of Covariates 2}. $[\hat{\mu}^1_{\mathcal{I}_1}( x_{i,1}=\{1,w_i\})](t)=[\hat{\mu}^2_{\mathcal{I}_1}( x_{i,2}=\{1,w_i\})](t)=\hat{\beta}_0(t) + \hat{\beta}_1(t) w_i$ 
\item \textit{Set of Covariates 3}. $[\hat{\mu}^1_{\mathcal{I}_1}( x_{i,1}=\{1,w_i,w_i^2\})](t)=[\hat{\mu}^2_{\mathcal{I}_1}( x_{i,2}=\{1,w_i,w_i^2\})](t)=\hat{\beta}_0(t) + \hat{\beta}_1(t) w_i + \hat{\beta}_2(t) w^2_i$ 
\end{itemize}
Note that `Set of Covariates 3' represents the case in which the model is correctly specified, while the other two sets of covariates represent a case of misspecification.

Table \ref{tab:validity_simstudy2}
\begin{table}[t]
\begin{center}
\centering
\begin{tabular}{|cccc|}
\hline
\multicolumn{1}{|c}{}&\multicolumn{3}{c|}{\cellcolor{light-gray}\textbf{Scenario 1}}\\
\cellcolor{light-int-gray}\textit{MPB method}&&&\\
\multicolumn{1}{|c}{}&\cellcolor{light-light-gray}Set of Cov. 1&\cellcolor{light-light-gray}Set of Cov. 2&\multicolumn{1}{c|}{\cellcolor{light-light-gray}Set of Cov. 3}\\ 
$\cellcolor{light-light-gray} n=20$&0.902[0.894,0.910]&0.897[0.889,0.906] &\multicolumn{1}{c|}{0.902[0.893,0.910]} \\  
$\cellcolor{light-light-gray}n=200$&0.907[0.899,0.915]&0.909[0.901,0.917] &\multicolumn{1}{c|}{0.899[0.890,0.907]} \\  
$\cellcolor{light-light-gray}n=2000$&0.899[0.891,0.908]&0.897[0.888,0.905] &\multicolumn{1}{c|}{0.904[0.896,0.913]} \\  
\cellcolor{light-int-gray}\textit{CUB method}&&&\\
\multicolumn{1}{|c}{}&\cellcolor{light-light-gray}Set of Cov. 1&\cellcolor{light-light-gray}Set of Cov. 2&\multicolumn{1}{c|}{\cellcolor{light-light-gray}Set of Cov. 3}\\ 
$\cellcolor{light-light-gray} n=20$&0.812[0.801,0.822]&0.811[0.800,0.822] &\multicolumn{1}{c|}{0.813[0.802,0.824]} \\  
$\cellcolor{light-light-gray}n=200$&0.820[0.809,0.830]&0.823[0.813,0.834] &\multicolumn{1}{c|}{0.805[0.794,0.816]} \\  
$\cellcolor{light-light-gray}n=2000$&0.808[0.797,0.819]&0.801[0.790,0.812] &\multicolumn{1}{c|}{0.815[0.804,0.826]} \\  \hline
\multicolumn{1}{|c}{}&\multicolumn{3}{c|}{\cellcolor{light-gray}\textbf{Scenario 2}}\\
\cellcolor{light-int-gray}\textit{MPB method}&&&\\
\multicolumn{1}{|c}{}&\cellcolor{light-light-gray}Set of Cov. 1&\cellcolor{light-light-gray}Set of Cov. 2&\multicolumn{1}{c|}{\cellcolor{light-light-gray}Set of Cov. 3}\\ 
$\cellcolor{light-light-gray} n=20$&0.899[0.890,0.907]&0.888[0.879,0.897] &\multicolumn{1}{c|}{0.905[0.894,0.915]} \\  
$\cellcolor{light-light-gray}n=200$&0.900[0.892,0.908]&0.895[0.887,0.904] &\multicolumn{1}{c|}{0.893[0.882,0.904]} \\  
$\cellcolor{light-light-gray}n=2000$&0.906[0.898,0.914]&0.899[0.891,0.908] &\multicolumn{1}{c|}{0.906[0.895,0.916]} \\  
\cellcolor{light-int-gray}\textit{CUB method}&&&\\
\multicolumn{1}{|c}{}&\cellcolor{light-light-gray}Set of Cov. 1&\cellcolor{light-light-gray}Set of Cov. 2&\multicolumn{1}{c|}{\cellcolor{light-light-gray}Set of Cov. 3}\\ 
$\cellcolor{light-light-gray} n=20$&0.838[0.828,0.848]&0.829[0.819,0.840] &\multicolumn{1}{c|}{0.844[0.834,0.854]} \\  
$\cellcolor{light-light-gray}n=200$&0.853[0.843,0.863]&0.843[0.833,0.853] &\multicolumn{1}{c|}{0.846[0.836,0.856]} \\  
$\cellcolor{light-light-gray}n=2000$&0.857[0.848,0.867]&0.852[0.843,0.862] &\multicolumn{1}{c|}{0.863[0.854,0.873]} \\  \hline
\end{tabular}

\caption{Simulation study 2: empirical coverage and related 95\% confidence interval in brackets for each combination of scenario, method, sample size and set of covariates. $\alpha=0.10$, set of modulation functions $\{ s^{\sigma}_{j,\mathcal{I}_1}\}_{j=1}^2$.}
\label{tab:validity_simstudy2}

\end{center}
\end{table}
shows the empirical coverage $\hat{p}$, together with the 95\% confidence interval defined as in Section \ref{sec::sim_study1}, obtained for each combination of scenario, method, sample size and set of covariates considering the set of modulation functions $\{ s^{\sigma}_{j,\mathcal{I}_1}\}_{j=1}^2$. In accordance with the results provided by the first simulation study, the MPB method presented in this article ensures empirical coverages very close to 0.90, with only two confidence intervals not including the target value $1-\alpha$. As regards the CUB method, in both scenarios the empirical coverages are far from $1-\alpha$ as expected, and  moving from Scenario 1 to Scenario 2 they grow due to the fact that the error terms are dependent. In particular, in the first scenario almost all the confidence intervals include the value $(1-\alpha)^2=0.81$, that is the coverage we expect from the CUB method since the error terms are independent. In view of this, the simulation study fully confirms the quite obvious conjecture that a carefully chosen multivariate approach must be considered in order to obtain proper multivariate simultaneous bands. 

Since the CUB method  does not guarantee the desired coverage, hereafter  we will only focus on the efficiency of the prediction bands outputted by the MPB method. Table \ref{tab:efficiency_simstudy2} 
\begin{table}[t]
\begin{center}
\centering
\begin{tabular}{|cccc|}
\hline
\multicolumn{1}{|c}{}&\multicolumn{3}{c|}{\cellcolor{light-gray}\textbf{Scenario 1}}\\
\multicolumn{1}{|c}{}&\cellcolor{light-light-gray}Set of Cov. 1&\cellcolor{light-light-gray}Set of Cov. 2&\multicolumn{1}{c|}{\cellcolor{light-light-gray}Set of Cov. 3}\\ 
$\cellcolor{light-light-gray} n=20$&5.606[4.940,6.539]&5.373[4.704,6.322] &\multicolumn{1}{c|}{6.007[5.106,7.212]} \\  
$\cellcolor{light-light-gray}n=200$&4.270[4.137,4.407]&3.820[3.706,3.943] &\multicolumn{1}{c|}{ 3.831[3.714,3.954]} \\  
$\cellcolor{light-light-gray}n=2000$&4.166[4.126,4.206]&3.707[3.673,3.741] &\multicolumn{1}{c|}{3.702[3.666,3.737]} \\   \hline
\multicolumn{1}{|c}{}&\multicolumn{3}{c|}{\cellcolor{light-gray}\textbf{Scenario 2}}\\
\multicolumn{1}{|c}{}&\cellcolor{light-light-gray}Set of Cov. 1&\cellcolor{light-light-gray}Set of Cov. 2&\multicolumn{1}{c|}{\cellcolor{light-light-gray}Set of Cov. 3}\\ 
$\cellcolor{light-light-gray} n=20$&5.279[4.623,6.186]&5.063[4.357,6.005] &\multicolumn{1}{c|}{5.637[4.788,6.814]} \\  
$\cellcolor{light-light-gray}n=200$&4.133[3.996,4.279]&3.690[3.571,3.815] &\multicolumn{1}{c|}{3.698[3.575,3.823]} \\  
$\cellcolor{light-light-gray}n=2000$&4.032[3.992,4.073]& 3.583 [3.547,3.618] &\multicolumn{1}{c|}{3.578[3.543,3.614]} \\  \hline
\multicolumn{1}{|c}{}&\multicolumn{3}{c|}{\cellcolor{light-gray}\textbf{Scenario 3}}\\
\multicolumn{1}{|c}{}&\cellcolor{light-light-gray}Set of Cov. 1&\cellcolor{light-light-gray}Set of Cov. 2&\multicolumn{1}{c|}{\cellcolor{light-light-gray}Set of Cov. 3}\\ 
$\cellcolor{light-light-gray} n=20$&4.820[4.184,5.649]&4.568[3.932,5.446] &\multicolumn{1}{c|}{5.052[4.273,6.141]} \\  
$\cellcolor{light-light-gray}n=200$&3.852[3.708,4.000]&3.442[3.319,3.565] &\multicolumn{1}{c|}{3.450[3.327,3.578]} \\  
$\cellcolor{light-light-gray}n=2000$&3.773[3.731,3.817]&3.354[3.315,3.391] &\multicolumn{1}{c|}{3.347[3.310,3.385]} \\  \hline

\end{tabular}

\caption{Simulation study 2: median size (first and third quartile in brackets) for each combination of scenario, sample size and set of covariates. MPB method, $\alpha=0.10$, set of modulation functions $\{ s^{\sigma}_{j,\mathcal{I}_1}\}_{j=1}^2$.}
\label{tab:efficiency_simstudy2}

\end{center}
\end{table}
shows the median size (together with the first and third quartile in brackets) of the prediction bands analyzed in Table \ref{tab:validity_simstudy2}. In addition to the two scenarios considered so far, Table  \ref{tab:efficiency_simstudy2} also analyzes a third scenario in which $ \varepsilon_{i1}(t)= \varepsilon_{i2}(t)$ $\forall i,t$ (and so $Y_{i1}=Y_{i2}$): despite its limited practical utility, this scenario represents an edge case that can provide useful information. First of all, for each combination of scenario and set of covariates, the size decreases when $n$ grows, both because improved regression estimates generally provide smaller nonconformity scores and because the value of $k^s$ is less dependent on random fluctuations. Focusing now on each combination of sample size and set of covariates, it is possible to notice that Scenario 1 typically  provides the biggest prediction bands, whereas Scenario 3 the smallest. From a practical point of view, this evidence is due to the nonconformity measure used: indeed, it searches the most extreme value of $ \left| (y_{dj}(t)-[\hat{\mu}^j_{\mathcal{I}_1}(x_{dj})](t))/s_{j,\mathcal{I}_1}(t)\right|$ over $\mathcal{T}_1, \mathcal{T}_2$, and so the nonconformity score computed when $y_{d1}=y_{d2}$ as in Scenario 3 will always be less than or equal to that computed when $y_{d1} \neq y_{d2}$. Scenario 2 represents an intermediate case between Scenario 1 and Scenario 3 as regards the structure of the error terms, and this is confirmed by the evidence provided by Table \ref{tab:efficiency_simstudy2}.

\subsection{Simulation Study 3: Efficiency}
\label{sec::sim_study3}
\begin{figure}
\begin{center}
\includegraphics[width=7.7625cm,height=6cm]{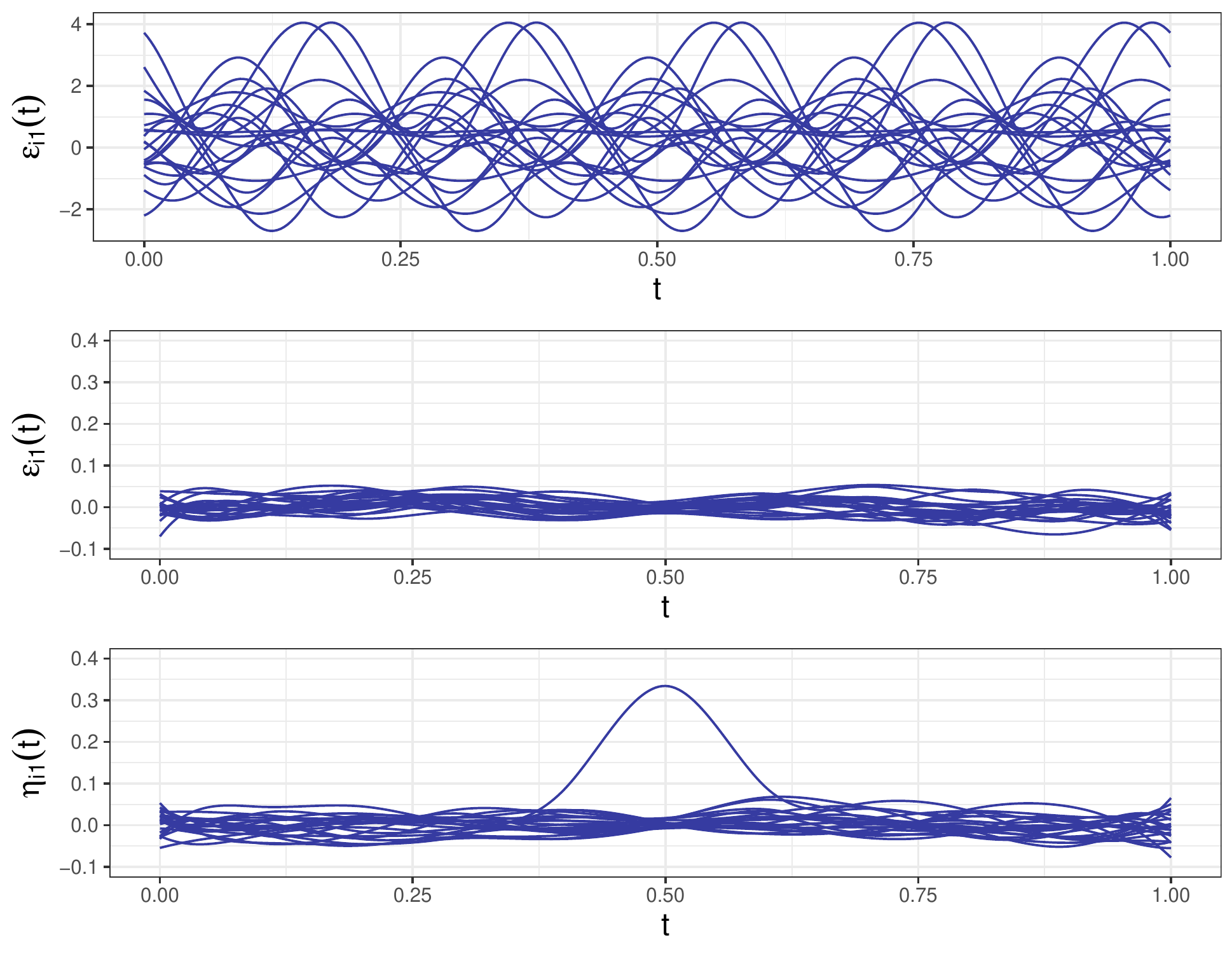} 
\end{center}
\caption{Example of realization of the error term related to $\{Y_{i1}\}_{i=1}^{n+1}$. First scenario at the top, second scenario in the middle, third scenario at the bottom. $n=20$.
 \label{fig::sim_study3} }
\end{figure}

The aim of the simulation study of this Section is to compare the three sets of modulation functions $(\{ s^{0}_{j}\}_{j=1}^p, \{ s^{\sigma}_{j,\mathcal{I}_1}\}_{j=1}^p, \{ \bar{s}_{j,\mathcal{I}_1}\}_{j=1}^p)$ in terms of efficiency. To do that, three different scenarios are taken into account: focusing just for now on the error terms and ignoring the systematic components, in the first scenario the error terms are characterized by a constant variability over the domains, in the second scenario the variability differs whereas in the third scenario the presence of outliers further complicates their specification. Formally, the three scenarios are:

\begin{itemize}
\item \textit{Scenario 1}. The two systematic components are defined as in the first scenario of Section \ref{sec::sim_study1}, while the independent functional errors $\varepsilon_{i1}(t), \varepsilon_{i2}(t)$ are defined as follows: 
\begin{align*}
\varepsilon_{i j}(t)=&B_{i+(n+1)(j-1), 1}+\\
                             &B_{i+(n+1)(j-1), 2} \cos \left(10 \pi\left(t+U_{i+(n+1)(j-1)}\right)\right)+ \\
                             &B_{i+(n+1)(j-1), 3} \sin \left(10 \pi\left(t+U_{i+(n+1)(j-1)}\right)\right) 
\end{align*}
$\forall i \in \{1,\dots,n+1\}, j \in \{1,2\}, t \in [0,1]$, with i.i.d. random vectors $\boldsymbol{B}_1,\dots,\boldsymbol{B}_{2(n+1)} \sim N_3(\boldsymbol{0},\Sigma)$, $\Sigma$ having the entries on the main diagonal equal to 1 and the entries outside the main diagonal equal to 0.7, i.i.d. random variables $U_1,\dots,U_{2(n+1)}  \sim U[-0.5,0.5]$.
\item \textit{Scenario 2}. The two systematic components are defined as in the first scenario of Section \ref{sec::sim_study1}, while the independent functional errors $\varepsilon_{i1}(t), \varepsilon_{i2}(t)$ are obtained by means of a B-spline basis expansion of order four, with 13 basis functions, equally spaced knots and normal random vectors as vectors of coefficients. Specifically, the $2\cdot(n+1)$ (observed) vectors of coefficients are independent realizations of $\boldsymbol{C}=(C_1,\dots,C_{13}) \sim N_{13}(\boldsymbol{0},\Sigma)$ with $\Sigma$ diagonal matrix such that $\mathrm{Var}[C_a]=0.001$ $\forall a \neq 7$,  $\mathrm{Var}[C_7]=9\cdot 10^{-6}$.
\item \textit{Scenario 3} 

\begin{align*}
Y_{i1}(t)=&\beta_0(t) +  \eta_{i1}(t), \quad i \in \{1,\dots,n+1\}, t \in [0,1] \\
Y_{i2}(t)=&\beta_0(t) +  \eta_{i2}(t), \quad i \in \{1,\dots,n+1\}, t \in [0,1] 
\end{align*}
with $\beta_0(t)=0$ $\forall t \in [0,1]$,
\begin{align*}
\eta_{ij}(t)=&\beta_1(t) w_{ij} +  \varepsilon_{ij}(t), \quad i \in \{1,\dots,n+1\}, j \in \{1,2\}, t \in [0,1] \\
\end{align*}
with $\beta_1(t)$ obtained by means of a B-spline basis expansion of order four, with 13 basis, equally spaced knots and all coefficients equal to 0 but the seventh equal to 0.5, $\varepsilon_{ij}(t)$ defined as in Scenario 2, and if $n=20$ then  $w_{ij}=0$ $\forall \{i,j\}\neq \{1,1\}$, $w_{1,1}=1$, whereas if $n \in \{200,2000\}$  then
\begin{equation*}
w_{ij}=\begin{cases}
1 & \text{if $i \in \big\{j+40\cdot \zeta: \zeta \in \{0,1,2,\dots,\frac{n}{40}-1 \} \big\}$}\\ 
0 & \text{otherwise}\\ 
\end{cases}
\end{equation*}
Despite the complex notation, the introduction of $w_{ij}$ is aimed at obtaining that $\sim 5\%$ of the multivariate functions $\boldsymbol{y}_1,\dots,\boldsymbol{y}_{n+1}$ (i.e. 1 out of 21 when $n=20$, 10 out of 201 when $n=200$, 100 out of 2001 when $n=2000$) is characterized, in one of the two components, by the anomalous behavior induced by $\beta_1(t)$. We propose such an unconventional structure for the error terms to simulate, for example, a regression framework in which relevant variables are not available. 
\end{itemize}
All three scenarios are evaluated considering only one set of covariates each, namely the case in which the corresponding model is correctly specified. Figure \ref{fig::sim_study3} 
shows, for each scenario, a realization of the error terms $\{\varepsilon_{i1}\}_{i=1}^{n+1}$ ($\{\eta_{i1}\}_{i=1}^{n+1}$ for Scenario 3) when $n=20$.

Table \ref{tab:efficiency_simstudy3} 
\begin{table}[t]
\begin{center}
\centering
\begin{tabular}{|ccccc c c c c|}
\hline
\multicolumn{1}{|c}{}&\multicolumn{3}{c|}{\cellcolor{light-gray}\textbf{Scenario 1}}\\
\multicolumn{1}{|c}{}&\cellcolor{light-light-gray}$\{ s^{0}_{j}\}_{j=1}^2$&\cellcolor{light-light-gray} $\{ s^{\sigma}_{j,\mathcal{I}_1}\}_{j=1}^2$ &\multicolumn{1}{c|}{\cellcolor{light-light-gray}$\{ \bar{s}_{j,\mathcal{I}_1}\}_{j=1}^2$}\\ 
$\cellcolor{light-light-gray} n=20$&9.599[8.348,11.116]&12.205[10.121,14.853] &\multicolumn{1}{c|}{14.241[11.730,17.798]} \\  
$\cellcolor{light-light-gray}n=200$&8.658[8.289,9.077]&8.835[8.442,9.246] &\multicolumn{1}{c|}{9.315[8.892,9.784]} \\  
$\cellcolor{light-light-gray}n=2000$&8.568[8.449,8.692]&8.587[8.469,8.712] &\multicolumn{1}{c|}{8.681[8.561,8.806]} \\  \hline
\multicolumn{1}{|c}{}&\multicolumn{3}{c|}{\cellcolor{light-gray}\textbf{Scenario 2}}\\
\multicolumn{1}{|c}{}&\cellcolor{light-light-gray}$\{ s^{0}_{j}\}_{j=1}^2$&\cellcolor{light-light-gray} $\{ s^{\sigma}_{j,\mathcal{I}_1}\}_{j=1}^2$ &\multicolumn{1}{c|}{\cellcolor{light-light-gray}$\{ \bar{s}_{j,\mathcal{I}_1}\}_{j=1}^2$}\\ 
$\cellcolor{light-light-gray}n=20$&0.168[0.152,0.188]&0.190[0.167,0.221] &\multicolumn{1}{c|}{0.213[0.186,0.249]} \\  
$\cellcolor{light-light-gray}n=200$&0.148[0.144,0.153]&0.126[0.123,0.130] &\multicolumn{1}{c|}{0.139[0.135,0.144]} \\  
$\cellcolor{light-light-gray}n=2000$&0.146[0.145,0.148]&0.122[0.121,0.123]&\multicolumn{1}{c|}{0.134[0.133,0.136]} \\  \hline
\multicolumn{1}{|c}{}&\multicolumn{3}{c|}{\cellcolor{light-gray}\textbf{Scenario 3}}\\
\multicolumn{1}{|c}{}&$\cellcolor{light-light-gray}\{s^{0}_{j}\}_{j=1}^2$& $\cellcolor{light-light-gray}\{ s^{\sigma}_{j,\mathcal{I}_1}\}_{j=1}^2$ &\multicolumn{1}{c|}{$\{\cellcolor{light-light-gray} \bar{s}_{j,\mathcal{I}_1}\}_{j=1}^2$}\\ 
$\cellcolor{light-light-gray}n=20$&0.201[0.157,0.667]&0.294[0.212,1.767] &\multicolumn{1}{c|}{0.407[0.277,1.869]} \\  
$\cellcolor{light-light-gray}n=200$&0.162[0.155,0.170]&0.167[0.161,0.172] &\multicolumn{1}{c|}{0.151[0.145,0.158]} \\  
$\cellcolor{light-light-gray}n=2000$&0.160[0.157,0.162]&0.161[0.160,0.163]&\multicolumn{1}{c|}{0.145[0.143,0.147]} \\  \hline
\end{tabular}

\caption{Simulation Study 3: median size (first and third quartile in brackets) for each combination of scenario, sample size and set of modulation functions. $\alpha=0.10$.}
\label{tab:efficiency_simstudy3}

\end{center}
\end{table}
shows the median size (defined as in Section \ref{sec::sim_study2}; first and third quartile in brackets) of the $N=5000$ prediction bands obtained for each combination of scenario, sample size and set of modulation functions. All three scenarios share the evidence that the prediction bands induced by $\{ s^{0}_{j}\}_{j=1}^2$ are typically smaller than those induced by the other two sets of modulation functions when the sample size is very small ($n=20$). This is due to the fact that regression estimates obtained with a small training set size likely provide an unreliable (and potentially misleading) set of modulation functions, leading to a preference for a set of modulation functions not depending on $\mathcal{I}_1$. As proof of that, it is not surprising that the two data-driven sets of modulation functions $\{ s^{\sigma}_{j,\mathcal{I}_1}\}_{j=1}^2, \{ \bar{s}_{j,\mathcal{I}_1}\}_{j=1}^2$ deliver the worst performance in the most complex Scenario, i.e. Scenario 3. Focusing on the other two sample sizes, in Scenario 1 the choice of not modulating seems appropriate due to the equal magnitude of the two components and the constant variability over $\mathcal{T}_1$, $\mathcal{T}_2$, but, as expected, the difference between the three alternative sets of modulation functions decreases when $n$ grows. Differently from Scenario 1, Scenario 2 is characterized by multivariate residuals showing a lower variability in the central portion of $\mathcal{T}_1$ and $\mathcal{T}_2$: as a consequence, $\{ s^{0}_{j}\}_{j=1}^2$ provides large prediction bands since it is not able to adapt the width of the band according to the local variability of the residuals, whereas $\{s^{\sigma}_{j,\mathcal{I}_1}\}_{j=1}^2$ is particularly effective since it induces a modulation process based on the two standard deviation functions. Finally, $\{ \bar{s}_{j,\mathcal{I}_1}\}_{j=1}^2$ represents the best solution in Scenario 3 given its ability to focus on the `least extreme' $\sim (1-\alpha)\cdot 100\%$ of data: indeed, differently from $\{ s^{0}_{j}\}_{j=1}^2$ it is able to reduce the width of the band in the central part of the domains, and differently from $\{s^{\sigma}_{j,\mathcal{I}_1}\}_{j=1}^2$ it does not uselessly enlarge the band in the same subinterval of $\mathcal{T}_1,\mathcal{T}_2$. Consequently, the simulation study seems to confirm the statistical intuition given in Section \ref{sec::choice_s} that the newly launched set of modulation functions $\{ \bar{s}_{j,\mathcal{I}_1}\}_{j=1}^p$ represents an interesting solution when functional residuals show a non-standard behavior and a modulation process driven by the value $1-\alpha$ is needed.

\section{Case Study: Analysis of Bike Mobility in the City of Milan}
\label{sec::bikemi}

\begin{figure}[t]
\begin{center}
\includegraphics[width=13cm,height=10.05cm]{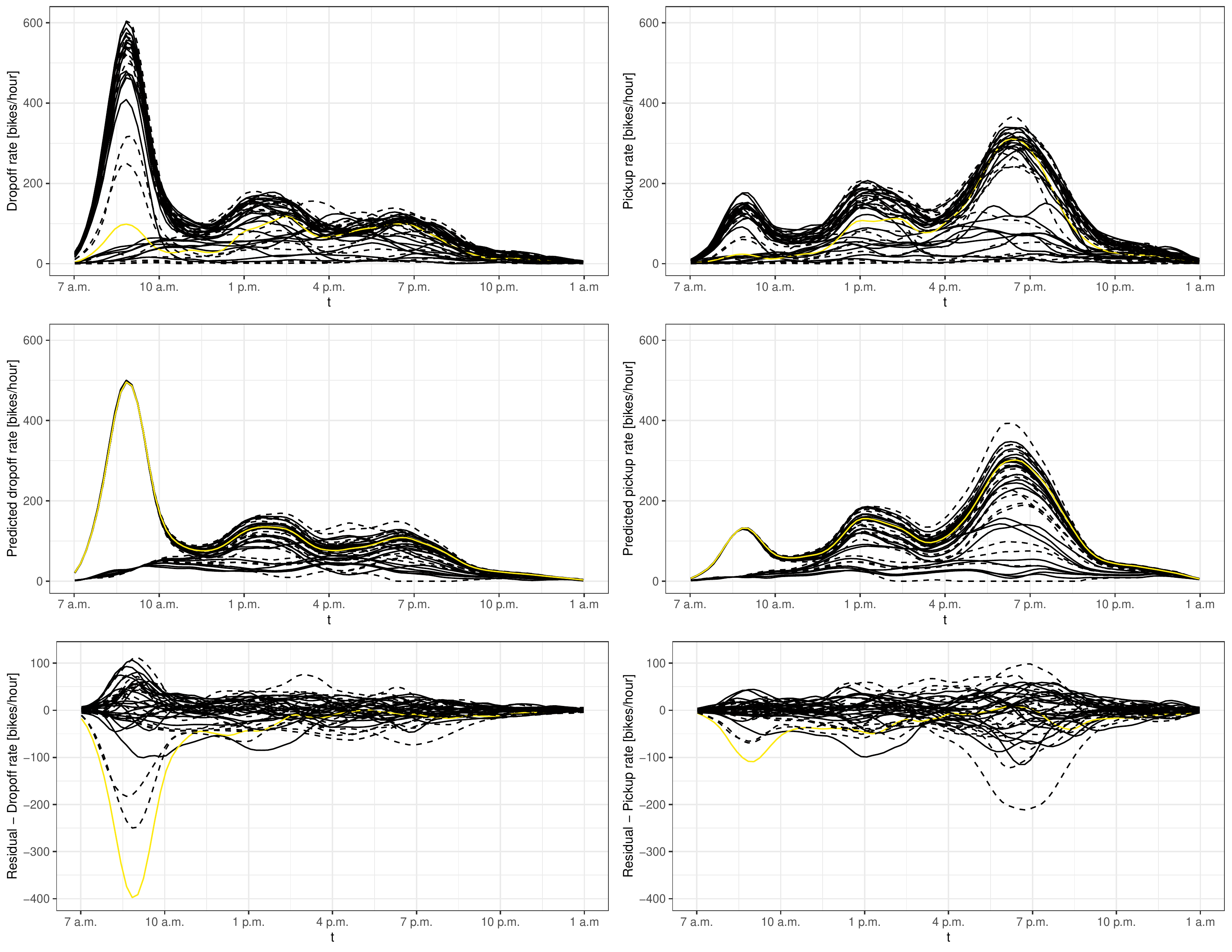} 
\end{center}
\caption{Dropoff and pickup rates (top left, top right respectively), corresponding functional predictions (center left, center right) and functional residuals (bottom left, bottom right). Yellow curves refer to 29 February; continuous curves refer to the observations in the training set, dashed curves to those in the calibration set. \label{fig::data_res_bikemi} }
\end{figure}

In order to illustrate the application potential of the method presented in this article, in this section we focus on a case study concerning urban mobility, and specifically the usage of a bike-sharing system in the Italian city of Milan. Moving from the raw data and the context presented in \citet{torti2021modelling}, the aim is to study the behavior of subscribers of Bikemi, a bike sharing system active in the city in which bikes are picked up and dropped off in specific docking stations located through the city. Starting from raw data providing various information about picked up bikes (simply \textit{pickups} hereafter) and dropped off bikes (simply \textit{dropoffs} hereafter) for each day considered, and focusing our attention - as an example - on the Duomo district only (i.e. the area in which Milan's cathedral is), the multivariate functional response variable $\boldsymbol{y}_i=(y_{i1},y_{i2})$ representing the rate of dropoffs ($y_{i1}$) and pickups ($y_{i2}$) is obtained via a standard kernel density estimation smoothing method \citep{hastie2009elements}.  In so doing, $y_{i1}(t)$ ($y_{i2}(t)$) represents the dropoff (pickup) rate at time t, with $t$ ranging from 7 a.m. day $i$ to 1 a.m. the next day (consequently, we assume that day $i$ ends at 1 a.m. the next day). The period considered starts on 25 January 2016 and ends on 6 March 2016: due to  an error in the data collection, 25 February is removed from the dataset in accordance with \citet{torti2021modelling}, and so the sample size is $n=41$. Data are shown in the two top panels  of Figure \ref{fig::data_res_bikemi}.

Like in \citet{torti2021modelling}, the regression estimates are obtained by fitting a concurrent functional-on-functional linear model \citep{ramsay_functional_2005}. The model hereby used includes as covariates a functional intercept, the temperature function (after subtracting the average daily temperature function of the period considered) in degrees Celsius, and a dummy variable indicating whether day $i$ is a weekday or not. Since the rates cannot be negative in any subinterval of the domain, the predicted functions are truncated to 0. However, as discussed in Section \ref{sec::NCM}, the purpose is to construct valid, meaningful and interpretable prediction bands also when simple regression estimators are specified, and so the choice of the covariates, as well as the functional form of the model, represents an aspect of limited interest in the framework considered.

\begin{figure}[t]
\begin{center}
\includegraphics[width=11.05cm,height=8.5425cm]{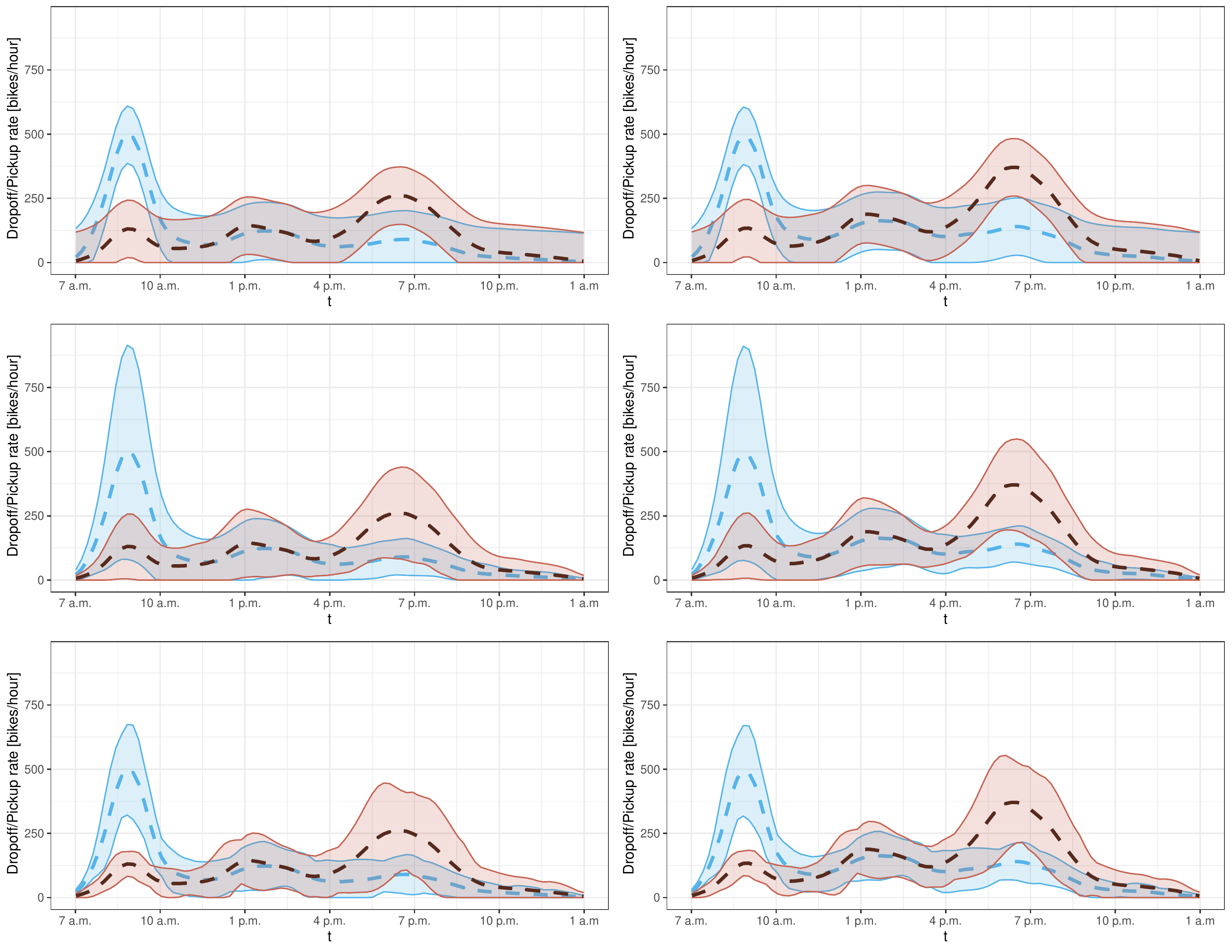} 
\end{center}
\caption{Prediction bands for the dropoff rate (light blue band) and the pickup rate (red band). Each panel refers to a combination of set of modulation functions ($\{ s^{0}_{j}\}_{j=1}^2$ at the top, $\{ s^{\sigma}_{j,\mathcal{I}_1}\}_{j=1}^2$ in the middle, $\{ \bar{s}_{j,\mathcal{I}_1}\}_{j=1}^2$ at the bottom) and scenario (first set of covariates on the left, second on the right). The  dashed lines indicate the corresponding regression estimates. $\alpha=0.25$. Split into calibration/training set: even/odd(+ day 20) days. \label{fig::pred_set_bikemi} }
\end{figure}

The method presented in Section \ref{sec::method} is performed by considering the three sets of modulation functions $\{ s^{0}_{j}\}_{j=1}^2, \{ s^{\sigma}_{j,\mathcal{I}_1}\}_{j=1}^2, \{ \bar{s}_{j,\mathcal{I}_1}\}_{j=1}^2$, $\alpha=0.25$ and $m=22$, $l=19$ in order to assign, as in the simulation studies, about half of the observations to the training set and to obtain the value $1-\lfloor (l+1)\alpha \rfloor/(l+1)$ equal to $1-\alpha$. To remain as neutral as possible, we will consider the case in which - after having labeled the days considered with numbers from 1 to 41 - the observations referring to an odd day are assigned to the training set and those referring to an even day to the calibration set, with the observation related to day 20 assigned to the training set to satisfy $m=22$. Two possible prediction scenarios are taken into account for the scope of visualization: in the first, we construct the multivariate prediction band for a weekday having the average temperature function of the period as temperature function; in the second, we construct it for a warmer than usual weekday (see Figure B.1 in Appendix B for a graphical representation of the two functional covariates, together with those observed). Figure \ref{fig::pred_set_bikemi} 
shows, for each of the three sets of modulation functions ($\{ s^{0}_{j}\}_{j=1}^2$ in the first row, $\{ s^{\sigma}_{j,\mathcal{I}_1}\}_{j=1}^2$ in the second row, $\{ \bar{s}_{j,\mathcal{I}_1}\}_{j=1}^2$ in the third row), the prediction sets induced by the two scenarios (first set of covariates in the first column, second in the second column). In particular, each panel shows the prediction band for the dropoff rate (light blue band) and the pickup rate (red band), with the two dashed lines representing the corresponding regression estimates. As for the predicted functions, the prediction bands are truncated to 0, as the rates cannot be negative in any subinterval of the domain. Note that this truncation does not involve any kind of drawback since the coverage reached by the prediction sets remains unchanged if a null probability portion of the bands is removed from the prediction bands. It is evident that the prediction bands for dropoffs induced by  $\{ s^{\sigma}_{j,\mathcal{I}_1}\}_{j=1}^2$ are quite large in the initial portion of the domain compared to those obtained by not modulating (i.e. $\{ s^{0}_{j}\}_{j=1}^2$) and by the proposed set $\{ \bar{s}_{j,\mathcal{I}_1}\}_{j=1}^2$. In order to clarify this aspect, let us consider Figure \ref{fig::data_res_bikemi}. Focusing on the  residual functions of the dropoff rates (i.e. the panel at the bottom left of the figure), it is easily noticeable that the yellow curve (referring to weekday 35, i.e. 29 February, which is assigned to the training set) shows an anomalous behavior in the initial part of the domain. The panel at the top left of the same figure suggests that this is due to the fact that day 35 was characterized by an unusually low dropoff rate compared to that observed in the other weekdays (which are the curves showing a pick around 9 a.m.). Consequently, by using $\{ s^{\sigma}_{j,\mathcal{I}_1}\}_{j=1}^2$ it is natural to obtain prediction bands for dropoffs extremely wide in the first portion of the domain since this outlier has a huge impact on the modulation process, while the corresponding prediction bands obtained by not modulating are not adversely affected as  $\{ s^{0}_{j}\}_{j=1}^2$ does not modulate the width of the band according to the local variability of the residuals. In view of this, the set of modulation functions $\{ \bar{s}_{j,\mathcal{I}_1}\}_{j=1}^2$ represents an intriguing solution since, in addition to modulate the width of the band along the domains, induces a modulation process which is not misled by the anomalous behavior of day 35. However, similar considerations would have been made also if other observations than the one related to day 35 had been assigned to the training set, as can be noticed by analyzing the functional residuals of the observations assigned to the calibration set in the two panels at the bottom of Figure \ref{fig::data_res_bikemi} (dashed curves for the calibration set; continuous curves for the training set). 
Despite the small sample size, the prediction sets of Figure \ref{fig::pred_set_bikemi} can provide profitable information: first of all, subscribers of Bikemi seem to mainly use bikes  to go to Duomo in the morning, whereas in the early evening the bike flow is reversed. Moving from the first set of covariates (weekday-temperature equal to the mean temperature of the period) to the second one (weekday-warm day), we notice that a higher temperature does not strongly affect people's behavior in the morning, whereas it involves a moderate increase in dropoffs and, at the same time, a big increase in pickups in the period of time around 7 p.m.. The information provided by the prediction bands can be indeed very useful to fleet managers in identifying the periods of time in which the imbalance between pickups and dropoffs could become critical based on the day of the week, the temperature function and other possible carefully chosen covariates.

\section{Conclusion and Further Developments}
\label{sec::conclusion}
In the present work we have developed a procedure aimed at creating prediction bands for multivariate functional data in a regression framework. Despite the paramount importance of this topic both from the methodological and applied point of view, to the best of our knowledge our method represents the first proposal in this direction. Moving from the approach proposed by \citet{diquigiovanni2021importance} for univariate i.i.d. functional data, the method presented in this article builds finite-sample either exact or valid prediction bands under the only assumption of exchangeable regression pairs with multivariate functional response. These properties, together with the fact that the procedure is scalable and the bands can be easily found in closed form, allow to obtain meaningful prediction bands regardless the regression estimator used, leading to a methodology which can be applied in a wide range of application scenarios. Moreover, we have introduced a specific set of modulation functions (namely $\{ \bar{s}_{j,\mathcal{I}_1}\}_{j=1}^p$) achieving an asymptotic result in terms of efficiency regardless the sample observed $\boldsymbol{z}_1,\dots,\boldsymbol{z}_n$ and inducing prediction bands whose width varies along the domains and across the components according to the local behavior. The simulation study and the real-world application provided in Section \ref{sec::sim_study} and \ref{sec::bikemi} respectively confirm the potential of the approach. Nevertheless, many possible directions still remain unexplored. Among these,  we plan to modify the methodology in order to apply it when regression data are dependent (as in the case, for example, of a functional time series); and, we plan to explore the impact of the regression estimator on the size of the prediction sets.

\section*{Supplementary material}

\subsection*{A Technical Proofs}
\label{sec::tech_proofs}
\subsubsection*{A.1 Proof of Section \ref{sec::NCM}}
\label{sec::proofs_31}

\textbf{\textit{Computation to find $\mathcal{C}^s_{n, 1-\alpha}(\boldsymbol{x}_{n+1})$}}

Since
\begin{align*}
\delta^s_{\boldsymbol{y}} =&  \frac{\left|\left\{d \in  \mathcal{I}_2 \cup \{n+1\} : R^s_{d} \geq R^s_{n+1}\right\}\right|}{l+1}, \\
\mathcal{C}^s_{n, 1-\alpha}\left(\boldsymbol{x}_{n+1}\right)=& \left\{\boldsymbol{y} \in  \prod_{j=1}^p L^{\infty}(\mathcal{T}_j): \delta^s_{\boldsymbol{y}}>\alpha \right\},
\end{align*}
if $\alpha \in [1/(l+1),1 )$, then $\boldsymbol{y} \in \mathcal{C}^s_{n, 1-\alpha}(\boldsymbol{x}_{n+1}) \iff R^s_{n+1} \leq k^s$, with $k^s$ the $\lceil (l+1)(1-\alpha) \rceil$th smallest value in the set $\{ R^s_d: d \in \mathcal{I}_2 \}$. Then 
\begin{align*}
&\sup_{j \in \{1,\dots,p\}} \left( \sup_{t \in \mathcal{T}_j} \left| \frac{y_{j}(t)-[\hat{\mu}^j_{\mathcal{I}_1}(x_{n+1,j})](t)}{s_{j,\mathcal{I}_1}(t)}\right| \right) \leq k^s \\
\iff  &\left| \frac{y_{j}(t)-[\hat{\mu}^j_{\mathcal{I}_1}(x_{n+1,j})](t)}{s_{j,\mathcal{I}_1}(t)}\right| \leq k^s \quad \forall j \in \{1,\dots,p\}, \forall t \in \mathcal{T}_j  \\
\iff &y_j(t) \in \big[ [\hat{\mu}^j_{\mathcal{I}_1}(x_{n+1,j})](t)-k^s \cdot s_{j,\mathcal{I}_1}(t), \\
& \hphantom{y_j(t) \in \big[} [\hat{\mu}^j_{\mathcal{I}_1}(x_{n+1,j})](t)+k^s \cdot s_{j,\mathcal{I}_1}(t)] \quad \forall j \in \{1,\dots,p\}, \forall t \in \mathcal{T}_j.
\end{align*}
As a consequence, the Split Conformal prediction set is
\begin{align*}
\mathcal{C}^s_{n, 1-\alpha}(\boldsymbol{x}_{n+1}):= \bigg\{ \boldsymbol{y} \in \prod_{j=1}^p L^{\infty}(\mathcal{T}_j):  y_j(t) \in \big[ &[\hat{\mu}^j_{\mathcal{I}_1}(x_{n+1,j})](t)-k^s \cdot s_{j,\mathcal{I}_1}(t), \\
&[\hat{\mu}^j_{\mathcal{I}_1}(x_{n+1,j})](t)+k^s \cdot s_{j,\mathcal{I}_1}(t)] \\
& \forall j \in \{1,\dots,p\}, \forall t \in \mathcal{T}_j \bigg\}.
\end{align*}

\textbf{\textit{Computation to find $\mathcal{C}^s_{n, 1-\alpha,\tau_{n+1}}(\boldsymbol{x}_{n+1})$}}

Consistently with the Split Conformal scenario, let us define
\begin{align*}
\delta^s_{\boldsymbol{y},\tau_{n+1}} :=& \frac{\left|\left\{d \in  \mathcal{I}_2: R^s_{d} > R^s_{n+1}\right\}\right| + \tau_{n+1} \left|\left\{d \in  \mathcal{I}_2 \cup \{n+1\}: R^s_{d} = R^s_{n+1}\right\}\right|}{l+1} \\
\mathcal{C}^s_{n, 1-\alpha,\tau_{n+1}}\left(\boldsymbol{x}_{n+1}\right):=& \left\{\boldsymbol{y} \in  \prod_{j=1}^p L^{\infty}(\mathcal{T}_j): \delta^s_{\boldsymbol{y},\tau_{n+1}}>\alpha \right\}.
\end{align*}

By definition, $\mathcal{C}^s_{n, 1-\alpha,1}(\boldsymbol{x}_{n+1})= \mathcal{C}^s_{n, 1-\alpha}(\boldsymbol{x}_{n+1})$.

Since $\delta_{\boldsymbol{y},\tau_{n+1}}^s \in [\tau_{n+1}/(l+1),  (l+\tau_{n+1})/(l+1)]$, we will focus on the scenario in which $\alpha \in [\tau_{n+1}/(l+1), (l+\tau_{n+1})/(l+1) )$. Let us define $w^s$ the $\lceil l + \tau_{n+1} - (l+1) \alpha  \rceil$th smallest value in the set $\{R_d: d \in \mathcal{I}_2 \}$, and $r^s_{n}$ ($v^s_n$ respectively) the number of elements in the set $\{R_d: d \in \mathcal{I}_2 \}$ that are equal to $w^s$ and that are to the right (left respectively) of $w^s$ in the sorted version of the set. Note that $r^s_{n}=v^s_{n}=0$ when the assumption about the continuous joint distribution of $\{ R_d : d \in \mathcal{I}_2\}$ is satisfied, but generally speaking we will consider $r^s_{n},v^s_{n} \in \mathcal{N}_{\geq0}$ such that $r^s_{n}+v^s_{n} \leq l-1$. By replicating calculations similar to those performed in the Split Conformal framework, we obtain that:
\begin{itemize}
\item if 
\begin{equation*}
\tau_{n+1}>\frac{(l+1)\alpha-\lfloor (l+1) \alpha -\tau_{n+1} \rfloor + r^s_{n}}{r^s_{n} + v^s_{n} + 2}
\end{equation*}
then $\boldsymbol{y} \in \mathcal{C}^s_{n, 1-\alpha,\tau_{n+1}}(\boldsymbol{x}_{n+1}) \iff R^s_{n+1} \leq w^s$ and so
\begin{align*}
\mathcal{C}^s_{n, 1-\alpha,\tau_{n+1}}(\boldsymbol{x}_{n+1}):= \bigg\{ \boldsymbol{y} \in \prod_{j=1}^p L^{\infty}(\mathcal{T}_j):  y_j(t) \in \big[ &[\hat{\mu}^j_{\mathcal{I}_1}(x_{n+1,j})](t)-w^s \cdot s_{j,\mathcal{I}_1}(t), \\
&[\hat{\mu}^j_{\mathcal{I}_1}(x_{n+1,j})](t)+w^s \cdot s_{j,\mathcal{I}_1}(t)] \\
& \forall j \in \{1,\dots,p\}, \forall t \in \mathcal{T}_j \bigg\}.
\end{align*}

\item if 
\begin{equation*}
\tau_{n+1} \leq \frac{(l+1)\alpha-\lfloor (l+1) \alpha -\tau_{n+1} \rfloor + r^s_{n}}{r^s_{n} + v^s_{n} + 2}
\end{equation*}
then $\boldsymbol{y} \in \mathcal{C}^s_{n, 1-\alpha,\tau_{n+1}}(\boldsymbol{x}_{n+1}) \iff R_{n+1} < w^s$ and so
\begin{align*}
\mathcal{C}^s_{n, 1-\alpha,\tau_{n+1}}(\boldsymbol{x}_{n+1}):= \bigg\{ \boldsymbol{y} \in \prod_{j=1}^p L^{\infty}(\mathcal{T}_j):  y_j(t) \in \big( &[\hat{\mu}^j_{\mathcal{I}_1}(x_{n+1,j})](t)-w^s \cdot s_{j,\mathcal{I}_1}(t), \\
&[\hat{\mu}^j_{\mathcal{I}_1}(x_{n+1,j})](t)+w^s \cdot s_{j,\mathcal{I}_1}(t)\big) \\
& \forall j \in \{1,\dots,p\}, \forall t \in \mathcal{T}_j \bigg\}.
\end{align*}
\end{itemize}

\textbf{\textit{Proof that the concatenation of the $p$ univariate prediction bands obtained by applying the nonconformity measure $\sup_{t \in \mathcal{T}_j} \left| \left( y_{j}(t)-[\hat{\mu}^j_{\mathcal{I}_1}(x_{j})](t) \right) / s_{j,\mathcal{I}_1}(t)\right| $ to the $p$ components separately is a subset of (\ref{eq:pred_set})  }}

Let us define $\mathcal{U}^s_{n, 1-\alpha}(\boldsymbol{x}_{n+1})$ as the multivariate prediction band obtained by concatenating the $p$ univariate prediction bands induced by applying the nonconformity measure $\sup_{t \in \mathcal{T}_j} \left| \left( y_{j}(t)-[\hat{\mu}^j_{\mathcal{I}_1}(x_{j})](t) \right) / s_{j,\mathcal{I}_1}(t)\right| $ to the $p$ components separately. Let us define, with a slight abuse of notation, 
\begin{equation*}
\tilde{R}^s_{dj}:=\sup_{t \in \mathcal{T}_j} \left| \left( y_{dj}(t)-[\hat{\mu}^j_{\mathcal{I}_1}(x_{dj})](t) \right) / s_{j,\mathcal{I}_1}(t)\right|, \quad \forall d \in \mathcal{I}_2, \forall j \in \{1,\dots,p\} \\
\end{equation*}
and $\tilde{k}^s_j$ the $\lceil (l+1)(1-\alpha) \rceil$th smallest value in the set $\{ \tilde{R}^s_{dj}: d \in \mathcal{I}_2 \}$.
By construction $R^s_d=\sup_{j \in \{1,\dots,p\}} \tilde{R}^s_{dj}$, and so $R^s_d \geq \tilde{R}^s_{dj}$ $\forall j \in \{1,\dots,p \}, d \in \mathcal{I}_2$ and then $k^s \geq \tilde{k}^s_j$ $\forall j \in \{ 1,\dots,p \}$.
In view of this, if $\boldsymbol{y} \in \mathcal{U}^s_{n, 1-\alpha}(\boldsymbol{x}_{n+1})$, i.e. 
\begin{align*}
y_j(t) \in \big[ &[\hat{\mu}^j_{\mathcal{I}_1}(x_{n+1,j})](t)-\tilde{k}^s_j \cdot s_{j,\mathcal{I}_1}(t),  \\ 
&[\hat{\mu}^j_{\mathcal{I}_1}(x_{n+1,j})](t)+\tilde{k}^s_j \cdot s_{j,\mathcal{I}_1}(t)] \quad \forall j \in \{1,\dots,p\}, \forall t \in \mathcal{T}_j, 
\end{align*}
then 
\begin{align*}
y_j(t) \in \big[ &[\hat{\mu}^j_{\mathcal{I}_1}(x_{n+1,j})](t)-k^s \cdot s_{j,\mathcal{I}_1}(t),  \\ 
&[\hat{\mu}^j_{\mathcal{I}_1}(x_{n+1,j})](t)+k^s \cdot s_{j,\mathcal{I}_1}(t)] \quad \forall j \in \{1,\dots,p\}, \forall t \in \mathcal{T}_j,
\end{align*}
i.e. $\boldsymbol{y} \in \mathcal{C}^s_{n, 1-\alpha}(\boldsymbol{x}_{n+1})$. 
As  $\boldsymbol{y} \in \mathcal{C}^s_{n, 1-\alpha}(\boldsymbol{x}_{n+1})$ does not necessarily imply $\boldsymbol{y} \in \mathcal{U}^s_{n, 1-\alpha}(\boldsymbol{x}_{n+1})$, then $\mathcal{U}^s_{n, 1-\alpha} \subseteq \mathcal{C}^s_{n, 1-\alpha}$.

\textbf{\textit{Proof that the concatenation of the pointwise prediction intervals obtained by applying the pointwise nonconformity measure $ \left| \left( y_{j}(t)-[\hat{\mu}^j_{\mathcal{I}_1}(x_{j})](t) \right) / s_{j,\mathcal{I}_1}(t)\right|$ $\forall j \in \{1,\dots,p\}$, $\forall t \in \mathcal{T}_j$ is a subset of (\ref{eq:pred_set})  }}

Let us define $\mathcal{U}^s_{n, 1-\alpha}(\boldsymbol{x}_{n+1})$ as the multivariate prediction band obtained by concatenating the pointwise prediction intervals obtained by applying the pointwise nonconformity measure $ \left| \left( y_{j}(t)-[\hat{\mu}^j_{\mathcal{I}_1}(x_{j})](t) \right) / s_{j,\mathcal{I}_1}(t)\right|$ $\forall j \in \{1,\dots,p\}$, $\forall t \in \mathcal{T}_j$. Let us define, with a slight abuse of notation, 
\begin{equation*}
\tilde{R}^s_{dj}(t):= \left| \left( y_{dj}(t)-[\hat{\mu}^j_{\mathcal{I}_1}(x_{dj})](t) \right) / s_{j,\mathcal{I}_1}(t)\right|, \quad \forall d \in \mathcal{I}_2, \forall j \in \{1,\dots,p\}, \forall t \in \mathcal{T}_j\\
\end{equation*}
and  $\tilde{k}^s_j(t)$ the $\lceil (l+1)(1-\alpha) \rceil$th smallest value in the set $\{ \tilde{R}^s_{dj}(t): d \in \mathcal{I}_2 \}$.
By construction $R^s_d=\sup_{j \in \{1,\dots,p\}} \left( \sup_{t \in \mathcal{T}_j} \tilde{R}^s_{dj}(t) \right)$, and so $R^s_d \geq \tilde{R}^s_{dj}(t)$ $\forall j \in \{1,\dots,p \}, d \in \mathcal{I}_2, t \in \mathcal{T}_j$ and then $k^s \geq \tilde{k}^s_j(t)$ $\forall j \in \{ 1,\dots,p \}, t \in \mathcal{T}_j$.
In view of this, if $\boldsymbol{y} \in \mathcal{U}^s_{n, 1-\alpha}(\boldsymbol{x}_{n+1})$, i.e. 
\begin{align*}
y_j(t) \in \big[ &[\hat{\mu}^j_{\mathcal{I}_1}(x_{n+1,j})](t)-\tilde{k}^s_j(t) \cdot s_{j,\mathcal{I}_1}(t),  \\ 
&[\hat{\mu}^j_{\mathcal{I}_1}(x_{n+1,j})](t)+\tilde{k}^s_j(t) \cdot s_{j,\mathcal{I}_1}(t)] \quad \forall j \in \{1,\dots,p\}, \forall t \in \mathcal{T}_j, 
\end{align*}
then 
\begin{align*}
y_j(t) \in \big[ &[\hat{\mu}^j_{\mathcal{I}_1}(x_{n+1,j})](t)-k^s \cdot s_{j,\mathcal{I}_1}(t),  \\ 
&[\hat{\mu}^j_{\mathcal{I}_1}(x_{n+1,j})](t)+k^s \cdot s_{j,\mathcal{I}_1}(t)] \quad \forall j \in \{1,\dots,p\}, \forall t \in \mathcal{T}_j,
\end{align*}
i.e. $\boldsymbol{y} \in \mathcal{C}^s_{n, 1-\alpha}(\boldsymbol{x}_{n+1})$. 
As  $\boldsymbol{y} \in \mathcal{C}^s_{n, 1-\alpha}(\boldsymbol{x}_{n+1})$ does not necessarily imply $\boldsymbol{y} \in \mathcal{U}^s_{n, 1-\alpha}(\boldsymbol{x}_{n+1})$, then $\mathcal{U}^s_{n, 1-\alpha} \subseteq \mathcal{C}^s_{n, 1-\alpha}$.

\textbf{\textit{Proof that prediction bands induced by $\{ s_{j,\mathcal{I}_1} \}_{j=1}^p$ and by $\{ \lambda \cdot s_{j,\mathcal{I}_1} \}_{j=1}^p$ coincide $\forall \lambda \in \mathbb{R}_{>0}$}}

Let $\mathcal{C}_{n, 1-\alpha}^{\lambda \cdot s}(\boldsymbol{x}_{n+1})$ be the prediction band induced by the set of modulation functions $\{ \lambda \cdot s_{j,\mathcal{I}_1} \}_{j=1}^p$. The nonconformity scores are:
\begin{align*}
R^{\lambda \cdot s}_d=&\sup_{j \in \{1,\dots,p\}} \left( \sup_{t \in \mathcal{T}_j} \left| \frac{y_{dj}(t)-[\hat{\mu}^j_{\mathcal{I}_1}(x_{dj})](t)}{\lambda \cdot s_{j,\mathcal{I}_1}(t)}\right| \right)=\frac{1}{\lambda} R^{s}_d,\quad d \in \mathcal{I}_2\\
R^{\lambda \cdot s}_{n+1}=& \sup_{j \in \{1,\dots,p\}} \left( \sup_{t \in \mathcal{T}_j} \left| \frac{y_{j}(t)-[\hat{\mu}^j_{\mathcal{I}_1}(x_{n+1,j})](t)}{\lambda \cdot s_{j,\mathcal{I}_1}(t)}\right| \right)=\frac{1}{\lambda} R^{s}_{n+1}.
\end{align*}
Moreover, let us define:
\begin{equation*}
\delta^{\lambda \cdot s}_{\boldsymbol{y}} :=  \frac{\left|\left\{d \in  \mathcal{I}_2 \cup \{n+1\} : R^{\lambda \cdot s}_{d} \geq R^{\lambda \cdot s}_{n+1}\right\}\right|}{l+1},
\end{equation*}
with, as usual, $\mathcal{C}_{n, 1-\alpha}^{\lambda \cdot s}(\boldsymbol{x}_{n+1}):= \left\{\boldsymbol{y} \in \prod_{j=1}^p L^{\infty}(\mathcal{T}_j): \delta^{\lambda \cdot s}_{\boldsymbol{y}}>\alpha \right\}$. As a consequence, 
$\boldsymbol{y} \in \mathcal{C}_{n, 1-\alpha}^{\lambda \cdot s}(\boldsymbol{x}_{n+1}) \iff R^{\lambda \cdot s}_{n+1} \leq k^{\lambda \cdot s}$, with $k^{\lambda \cdot s}$ the $\lceil (l+1)(1-\alpha) \rceil$th smallest value in the set $\{ R^{\lambda \cdot s}_d: d \in \mathcal{I}_2 \}$. Since $R^{\lambda \cdot s}_d= R^{s}_d/ \lambda$ $\forall d \in \mathcal{I}_2 $, then $k^{\lambda \cdot s}= k^s/\lambda$. Then:

\begin{align*}
&\phantom{\Rightarrow} R^{\lambda \cdot s}_{n+1} \leq k^{\lambda \cdot s} \\
&\iff \frac{1}{\lambda} R^{s}_{n+1} \leq \frac{k^s}{\lambda}  \\
&\iff R^{s}_{n+1} \leq k^s,  \\
\end{align*}

and since $\boldsymbol{y} \in \mathcal{C}_{n, 1-\alpha}^{s}(\boldsymbol{x}_{n+1}) \iff R^{s}_{n+1} \leq k^s$, then $\mathcal{C}_{n, 1-\alpha}^{\lambda \cdot s}(\boldsymbol{x}_{n+1})$ coincides with $\mathcal{C}_{n, 1-\alpha}^{s}(\boldsymbol{x}_{n+1})$.

\subsubsection*{A.2 Proof of Section \ref{sec::choice_s}}
\label{sec::proofs_32}

\textbf{\textit{Proof of Theorem \ref{th:shared_convergence}}}

Let us consider $\bar{s}_{j, \mathcal{I}_1}(t)$, with $j \in \{1,\dots,p \}$.
Since $m/n=\theta$ with $0 < \theta < 1$, if $n \to +\infty$ then $m \to +\infty$. The scalar $\gamma$ is the empirical quantile of order $\lceil (m+1)(1-\alpha) \rceil)$ of $\{ \sup_{j \in \left\{1,\dots,p\right\}} \left( \sup_{t \in \mathcal{T}_j} \left| y_{hj}(t)-[\hat{\mu}^j_{\mathcal{I}_1}(x_{hj})](t) \right| \right) : h \in \mathcal{I}_1 \}$. First of all note that 
\begin{align*}
\lim_{m \to +\infty} \frac{\lceil (m+1)(1-\alpha) \rceil}{m}= \lim_{m \to +\infty} \frac{m+1 - \lfloor (m+1)\alpha \rfloor}{m}
\end{align*}
and since
\begin{equation*}
\frac{(m+1)\alpha-1}{m} \leq \frac{\lfloor(m+1)\alpha \rfloor}{m} \leq \frac{(m+1)\alpha}{m} \quad \forall m \in  \mathbb{N},
\end{equation*}
\begin{equation*}
\lim_{m \to +\infty} \frac{(m+1)\alpha-1}{m} = \lim_{m \to +\infty} \frac{(m+1)\alpha}{m} = \alpha
\end{equation*}

then by the squeeze theorem we know that
\begin{equation*}
\lim_{m \to +\infty} \frac{\lfloor(m+1)\alpha \rfloor}{m} = \alpha
\end{equation*}
and then
\begin{align*}
\lim_{m \to +\infty} \frac{\lceil (m+1)(1-\alpha) \rceil)}{m}= 1-\alpha.
\end{align*}

Consequently, $\gamma$ is the empirical quantile of order  $1-\alpha$ when $m \rightarrow +\infty$.

Let us define $w_h:=\sup_{j \in \left\{1,\dots,p\right\}} \left( \sup_{t \in \mathcal{T}_j} \left| y_{hj}(t)-[\hat{\mu}^j_{\mathcal{I}_1}(x_{hj})](t) \right| \right)$ $\forall$ $h \in \mathcal{I}_1$. The random variables $\{W_h: h \in \mathcal{I}_1 \}$ from which $\{w_h: h \in \mathcal{I}_1 \}$ are drawn are continuous and after $\mathrm{Var}\big[[\hat{\mu}^j_{\mathcal{I}_1}(X_{hj})](t)\big] \to 0$ $\forall j \in \{1,\dots,p\}$ they become i.i.d.. The Glivenko-Cantelli theorem guarantees that the empirical distribution function of these variables converges uniformly and almost surely pointwise to its distribution function, and so also the empirical quantiles converge in distribution - and so in probability - to the corresponding theoretical quantiles \citep[see, for example,][ chap. 21]{van2000asymptotic}. In so doing, empirical quantile $\gamma$ converges to $q_{1-\alpha}$, the theoretical quantile of order $1-\alpha$. As a consequence:
\begin{equation*}
\mathcal{H}_1:=\{h \in \mathcal{I}_1: \sup_{j \in \left\{1,\dots,p\right\}} \left( \sup_{t \in \mathcal{T}_j} \left| y_{hj}(t)-[\hat{\mu}^j_{\mathcal{I}_1}(x_{hj})](t) \right| \right) \leq q_{1-\alpha}\} 
\end{equation*}
when $m \to +\infty$, with $q_{1-\alpha}$ non-random quantity.  Let us consider the numerator of $\bar{s}_{j,\mathcal{I}_1}(t)$ $\forall j \in \{1,\dots,p\}$ 
as the denominator is a normalizing constant. $\forall t \in \mathcal{T}_j$, the sequence $\{\max_{h \in \mathcal{H}_1}|y_{hj}(t)-[\hat{\mu}^j_{\mathcal{I}_1}(x_{hj})](t) |\}_{m}$ is eventually bounded by $q_{1-\alpha}$ and is eventually increasing since $\{\vert \mathcal{H}_1 \vert\}_m$ is eventually increasing. Therefore  the sequence converges to its supremum by the monotone convergence theorem.

As regards $\bar{s}^{c}_{j,\mathcal{I}_1,\mathcal{I}_2}$, first of all it is possible to notice that  if $n \to +\infty$ then $l=n(1-\theta) \to +\infty$. In order to show the convergence of the numerator of $\bar{s}^{c}_{j,\mathcal{I}_1,\mathcal{I}_2}$ to the same limit function,   it is sufficient to consider the previous calculations by substituting  $\gamma$ with $k$, $m$ with $l$, $\mathcal{H}_1$ with $\mathcal{H}_2$ and $\mathcal{I}_1$ with $\mathcal{I}_2$ (except for $[\hat{\mu}^j_{\mathcal{I}_1}(x_{hj})](t)$ that is not substituted by $[\hat{\mu}^j_{\mathcal{I}_2}(x_{hj})](t)$). Finally, as the numerators of $\bar{s}_{j,\mathcal{I}_1,\mathcal{I}_2}$ and $\bar{s}^{c}_{j,\mathcal{I}_1}$ converge to the same function $\forall j \in \{1,\dots,p\}$, also the two normalizing constants converge to the same value.

\textbf{\textit{Proof of Theorem \ref{th:better_than_not_modul}}}



For the sake of simplicity, let us focus on the case in which $\vert \mathcal{H}_2 \vert = \lceil (l+1)(1-\alpha) \rceil $. Under the assumption concerning the continuous joint distribution of $\{ R_d : d \in \mathcal{I}_2\}$ made in Secton \ref{sec::conf_pred}  such condition is always satisfied, but for the sake of completeness the proof when this assumption is violated is addressed below.

\begin{itemize}
\item $\forall d \in \mathcal{H}_2$, $\forall j \in \{1,\dots,p\}$ the following relationship holds $\forall t \in \mathcal{T}_j$:
\begin{align*}
\phantom{=}& \left| \frac{y_{dj}(t)- [\hat{\mu}^j_{\mathcal{I}_1}(x_{dj})](t) }{\bar{s}^{c}_{j,\mathcal{I}_1,\mathcal{I}_2}(t)} \right| \\
=& \sum_{j=1}^p \int_{\mathcal{T}_j}\max_{d \in \mathcal{H}_2} |y_{dj}(t)-[\hat{\mu}^j_{\mathcal{I}_1}(x_{dj})](t) | dt  \cdot   \frac{\left| y_{dj}(t)- [\hat{\mu}^j_{\mathcal{I}_1}(x_{dj})](t) \right|}{\max_{d \in \mathcal{H}_2} |y_{dj}(t)-[\hat{\mu}^j_{\mathcal{I}_1}(x_{dj})](t) |}  \\
\leq& \sum_{j=1}^p \int_{\mathcal{T}_j}\max_{d \in \mathcal{H}_2} |y_{dj}(t)-[\hat{\mu}^j_{\mathcal{I}_1}(x_{dj})](t) | dt,
\end{align*}
and then
\begin{align*}
R^{\bar{s}^{c}}_d:=  \sup_{j \in \left\{1,\dots,p\right\}} \left( \sup_{t \in \mathcal{T}_j} \left| \frac{y_{dj}(t)- [\hat{\mu}^j_{\mathcal{I}_1}(x_{dj})](t) }{\bar{s}^{c}_{j,\mathcal{I}_1,\mathcal{I}_2}(t)} \right| \right) \leq \sum_{j=1}^p \int_{\mathcal{T}_j}\max_{d \in \mathcal{H}_2} |y_{dj}(t)-[\hat{\mu}^j_{\mathcal{I}_1}(x_{dj})](t) | dt.
\end{align*}

Specifically, $\exists$ $\underline{d} \in  \mathcal{H}_2$ such that $R^{\bar{s}^{c}}_{\underline{d}}= \sum_{j=1}^p \int_{\mathcal{T}_j}\max_{d \in \mathcal{H}_2} |y_{dj}(t)-[\hat{\mu}^j_{\mathcal{I}_1}(x_{dj})](t) | dt $ since $\forall j \in \{1,\dots,p\}$ and $\forall t \in \mathcal{T}_j$ at least one function $y_{\underline{d}, j}$ satisfies $\left| y_{\underline{d}j}(t)- [\hat{\mu}^j_{\mathcal{I}_1}(x_{\underline{d}j})](t) \right|=\max_{d \in \mathcal{H}_2} |y_{dj}(t)-[\hat{\mu}^j_{\mathcal{I}_1}(x_{dj})](t) |$.

\item Let us define  $\mathcal{CH}_2:=\mathcal{I}_2 \setminus \mathcal{H}_2$  and let $(t^{*}_d, j^{*}_d)$ be the couple of values such that 
\begin{equation*} 
 \left| y_{d{j^{*}_d}}(t^{*}_d)- [\hat{\mu}^{j^{*}_d}_{\mathcal{I}_1}(x_{d{j^{*}_d}})](t^{*}_d)  \right|=\sup_{j \in \left\{1,\dots,p\right\}} \left( \sup_{t \in \mathcal{T}_j} \left| y_{dj}(t)-[\hat{\mu}^j_{\mathcal{I}_1}(x_{dj})](t) \right| \right) \quad \forall d \in \mathcal{I}_2.
\end{equation*} 

If $(t^{*}_d, b^{*}_d)$ is not unique, it is randomly chosen from the couples satisfying that condition. 

$\forall b \in \mathcal{CH}_2$,  by definition of $\mathcal{H}_2$ it is possible to notice that $ \left| y_{b{j^{*}_b}}(t^{*}_b)- [\hat{\mu}^{j^{*}_b}_{\mathcal{I}_1}(x_{b{j^{*}_b}})](t^{*}_b)  \right| >\max_{d \in \mathcal{H}_2} |y_{d{j^{*}_b}}(t^{*}_b)-[\hat{\mu}^{j^{*}_b}_{\mathcal{I}_1}(x_{d{j^{*}_b}})](t^{*}_b) |$ and so  the following relationship holds:
\begin{align*}
\phantom{=}& \left| \frac{y_{b{j^{*}_b}}(t^{*}_b)- [\hat{\mu}^{j^{*}_b}_{\mathcal{I}_1}(x_{b{j^{*}_b}})](t^{*}_b)}{\bar{s}^{c}_{j^{*}_b, \mathcal{I}_1,\mathcal{I}_2}(t^{*}_b)} \right| \\
=& \sum_{j=1}^p \int_{\mathcal{T}_j}\max_{d \in \mathcal{H}_2} |y_{dj}(t)-[\hat{\mu}^j_{\mathcal{I}_1}(x_{dj})](t) | dt  \cdot   \frac{\left|y_{b{j^{*}_b}}(t^{*}_b)- [\hat{\mu}^{j^{*}_b}_{\mathcal{I}_1}(x_{b{j^{*}_b}})](t^{*}_b) \right|}{\max_{d \in \mathcal{H}_2} |y_{d{j^{*}_b}}(t^{*}_b)-[\hat{\mu}^{j^{*}_b}_{\mathcal{I}_1}(x_{d{j^{*}_b}})](t^{*}_b) |}  \\
>&  \sum_{j=1}^p \int_{\mathcal{T}_j}\max_{d \in \mathcal{H}_2} |y_{dj}(t)-[\hat{\mu}^j_{\mathcal{I}_1}(x_{dj})](t) | dt.
\end{align*}
Consequently, 
\begin{align*}
R^{\bar{s}^{c}}_b:=  \sup_{j \in \left\{1,\dots,p\right\}} \left( \sup_{t \in \mathcal{T}_j} \left| \frac{y_{bj}(t)- [\hat{\mu}^j_{\mathcal{I}_1}(x_{bj})](t) }{\bar{s}^{c}_{j,\mathcal{I}_1,\mathcal{I}_2}(t)} \right| \right) > \sum_{j=1}^p \int_{\mathcal{T}_j}\max_{d \in \mathcal{H}_2} |y_{dj}(t)-[\hat{\mu}^j_{\mathcal{I}_1}(x_{dj})](t) | dt.
\end{align*}
\end{itemize}

Since:
\begin{itemize}
\item $\vert \mathcal{H}_2 \vert = \lceil (l+1)(1-\alpha) \rceil $
\item $\forall d \in \mathcal{H}_2$ $R^{\bar{s}^{c}}_d \leq  \sum_{j=1}^p \int_{\mathcal{T}_j}\max_{d \in \mathcal{H}_2} |y_{dj}(t)-[\hat{\mu}^j_{\mathcal{I}_1}(x_{dj})](t) | dt$ and  $\exists$ $\underline{d} \in  \mathcal{H}_2$ such that $R^{\bar{s}^{c}}_{\underline{d}}=  \sum_{j=1}^p \int_{\mathcal{T}_j}\max_{d \in \mathcal{H}_2} |y_{dj}(t)-[\hat{\mu}^j_{\mathcal{I}_1}(x_{dj})](t) | dt$
\item $\forall b \in \mathcal{CH}_2$ $R^{\bar{s}^{c}}_b > \sum_{j=1}^p \int_{\mathcal{T}_j}\max_{d \in \mathcal{H}_2} |y_{dj}(t)-[\hat{\mu}^j_{\mathcal{I}_1}(x_{dj})](t) | dt$
\end{itemize}

we conclude that $k^{\bar{s}^{c}}= \sum_{j=1}^p \int_{\mathcal{T}_j}\max_{d \in \mathcal{H}_2} |y_{dj}(t)-[\hat{\mu}^j_{\mathcal{I}_1}(x_{dj})](t) | dt$, with $k^{\bar{s}^{c}}$ the $\lceil (l+1)(1-\alpha) \rceil$th smallest value in the set $\{R^{\bar{s}^{c}}_d: d \in \mathcal{I}_2 \}$.

If $\vert \mathcal{H}_2 \vert > \lceil (l+1)(1-\alpha) \rceil $, then $R^{\bar{s}^{c}}_{d}=  \sum_{j=1}^p \int_{\mathcal{T}_j}\max_{d \in \mathcal{H}_2} |y_{dj}(t)-[\hat{\mu}^j_{\mathcal{I}_1}(x_{dj})](t) | dt $ is valid $\forall d \in \mathcal{H}_2$ such that $\sup_{j \in \left\{1,\dots,p\right\}} \left( \sup_{t \in \mathcal{T}_j} |y_{dj}(t)-[\hat{\mu}^j_{\mathcal{I}_1}(x_{dj})](t) | \right)=k$ 
and  we can conclude also in this case that $k^{\bar{s}^{c}}= \sum_{j=1}^p \int_{\mathcal{T}_j}\max_{d \in \mathcal{H}_2} |y_{dj}(t)-[\hat{\mu}^j_{\mathcal{I}_1}(x_{dj})](t) | dt$.

Focusing now on the set of modulation functions $s^0$, $\forall d \in \mathcal{I}_2$:
\begin{equation*}
R^{s^{0}}_d:=  \sup_{j \in \left\{1,\dots,p\right\}} \left( \sup_{t \in \mathcal{T}_j} \left| \frac{y_{dj}(t)- [\hat{\mu}^j_{\mathcal{I}_1}(x_{dj})](t) }{s^{0}_{j}(t)} \right| \right) =  \sup_{j \in \left\{1,\dots,p\right\}} \left( \sup_{t \in \mathcal{T}_j} \left| y_{dj}(t)- [\hat{\mu}^j_{\mathcal{I}_1}(x_{dj})](t) \right| \right) \cdot   \sum_{j=1}^p  \left| \mathcal{T}_j \right|.
\end{equation*}
Since $k^{s^{0}}$ is the $\lceil (l+1)(1-\alpha) \rceil$th smallest value in the set $\{R^{s^{0}}_d: d \in \mathcal{I}_2 \}$, by definition of $\mathcal{H}_2$ we can notice that
\begin{align*}
k^{s^{0}} =& \max_{d \in \mathcal{H}_2} R^{s^{0}}_d \\
=& \max_{d \in \mathcal{H}_2} \left( \sup_{j \in \left\{1,\dots,p\right\}} \left( \sup_{t \in \mathcal{T}_j} \left| y_{dj}(t)- [\hat{\mu}^j_{\mathcal{I}_1}(x_{dj})](t) \right| \right) \right) \cdot   \sum_{j=1}^p  \left| \mathcal{T}_j \right| \\
=& \sup_{j \in \left\{1,\dots,p\right\}}  \left( \sup_{t \in \mathcal{T}_j} \left( \max_{d \in \mathcal{H}_2} \left| y_{dj}(t)- [\hat{\mu}^j_{\mathcal{I}_1}(x_{dj})](t) \right| \right) \right) \cdot   \sum_{j=1}^p  \left| \mathcal{T}_j \right| .
\end{align*}

Since by the integral mean value theorem we know that $\forall j \in \{1,\dots,p\}$
\begin{equation*}
\sup_{t \in \mathcal{T}_j} \left( \max_{d \in \mathcal{H}_2} \left| y_{dj}(t)- [\hat{\mu}^j_{\mathcal{I}_1}(x_{dj})](t) \right| \right) \cdot \left| \mathcal{T}_j \right|\geq \int_{\mathcal{T}_j} \max_{d \in \mathcal{H}_2} \left| y_{dj}(t)- [\hat{\mu}^j_{\mathcal{I}_1}(x_{dj})](t) \right| dt,
\end{equation*}

then the following relationship is valid:
\begin{equation}
\sum_{j=1}^p  \sup_{t \in \mathcal{T}_j} \left( \max_{d \in \mathcal{H}_2} \left| y_{dj}(t)- [\hat{\mu}^j_{\mathcal{I}_1}(x_{dj})](t) \right| \right) \cdot \left| \mathcal{T}_j \right| \geq \sum_{j=1}^p \int_{\mathcal{T}_j} \max_{d \in \mathcal{H}_2} \left| y_{dj}(t)- [\hat{\mu}^j_{\mathcal{I}_1}(x_{dj})](t) \right| dt.
\label{dis_fun_1}
\end{equation}

In addition, by definition $\forall j \in \{1,\dots,p\}$
\begin{equation*}
\sup_{j \in \left\{1,\dots,p\right\}} \left( \sup_{t \in \mathcal{T}_j} \left( \max_{d \in \mathcal{H}_2} \left| y_{dj}(t)- [\hat{\mu}^j_{\mathcal{I}_1}(x_{dj})](t) \right| \right) \right) \geq  \sup_{t \in \mathcal{T}_j} \left( \max_{d \in \mathcal{H}_2} \left| y_{dj}(t)- [\hat{\mu}^j_{\mathcal{I}_1}(x_{dj})](t) \right| \right)
\end{equation*}

and so: 

\begin{align}
 &  \sum_{j=1}^p   \sup_{j \in \left\{1,\dots,p\right\}}  \left( \sup_{t \in \mathcal{T}_j} \left( \max_{d \in \mathcal{H}_2} \left| y_{dj}(t)- [\hat{\mu}^j_{\mathcal{I}_1}(x_{dj})](t) \right| \right) \right)  \cdot  \left| \mathcal{T}_j \right| \nonumber \\
=& \sup_{j \in \left\{1,\dots,p\right\}}  \left( \sup_{t \in \mathcal{T}_j} \left( \max_{d \in \mathcal{H}_2} \left| y_{dj}(t)- [\hat{\mu}^j_{\mathcal{I}_1}(x_{dj})](t) \right| \right) \right) \cdot   \sum_{j=1}^p  \left| \mathcal{T}_j \right|  \nonumber \\
\geq& \sum_{j=1}^p  \sup_{t \in \mathcal{T}_j} \left( \max_{d \in \mathcal{H}_2} \left| y_{dj}(t)- [\hat{\mu}^j_{\mathcal{I}_1}(x_{dj})](t) \right| \right) \cdot \left| \mathcal{T}_j \right|. \nonumber \\
\label{dis_fun_2} 
\end{align}

By combining (\ref{dis_fun_1}) and (\ref{dis_fun_2})  we can notice that 
\begin{equation*}
\sup_{j \in \left\{1,\dots,p\right\}}  \left( \sup_{t \in \mathcal{T}_j} \left( \max_{d \in \mathcal{H}_2} \left| y_{dj}(t)- [\hat{\mu}^j_{\mathcal{I}_1}(x_{dj})](t) \right| \right) \right) \cdot   \sum_{j=1}^p  \left| \mathcal{T}_j \right|  \geq  \sum_{j=1}^p \int_{\mathcal{T}_j} \max_{d \in \mathcal{H}_2} \left| y_{dj}(t)- [\hat{\mu}^j_{\mathcal{I}_1}(x_{dj})](t) \right| dt,
\end{equation*}
i.e. $k^{s^{0}} \geq k^{\bar{s}^{c}}$. Then, $\mathcal{Q}(s^{0}) \geq \mathcal{Q}( \bar{s}^{c}_{\mathcal{I}_1,\mathcal{I}_2})$.

Specifically, the integral mean value theorem guarantees that $\forall j \in \{1,\dots,p\}$
\begin{align*}
\sup_{t \in \mathcal{T}_j} \left( \max_{d \in \mathcal{H}_2} \left| y_{dj}(t)- [\hat{\mu}^j_{\mathcal{I}_1}(x_{dj})](t) \right| \right) \cdot \left| \mathcal{T}_j \right| = \int_{\mathcal{T}_j} \max_{d \in \mathcal{H}_2} \left| y_{dj}(t)- [\hat{\mu}^j_{\mathcal{I}_1}(x_{dj})](t) \right| dt \\
\iff  \max_{d \in \mathcal{H}_2} \left| y_{dj}(t)- [\hat{\mu}^j_{\mathcal{I}_1}(x_{dj})](t) \right| \quad \text{is constant almost everywhere},
\end{align*}

i.e. if and only if $  \bar{s}^{c}_{j,\mathcal{I}_1,\mathcal{I}_2}(t)$ is constant almost everywhere over $\mathcal{T}_j$. Consequently, if  at least one of the functions $\bar{s}^c_{1, \mathcal{I}_1,\mathcal{I}_2}(t), \dots, \bar{s}^c_{p, \mathcal{I}_1,\mathcal{I}_2}(t)$ is not constant almost everywhere over its domain then the left side of (\ref{dis_fun_1}) is strictly greater than  the right side (implying $\mathcal{Q}(s^{0}) > \mathcal{Q}( \bar{s}^{c}_{\mathcal{I}_1,\mathcal{I}_2})$); otherwise, $\mathcal{Q}(s^{0}) = \mathcal{Q}( \bar{s}^{c}_{\mathcal{I}_1,\mathcal{I}_2})$.

\textbf{\textit{Generalization of $(\bar{s}_{\mathcal{I}_1}, \bar{s}^{c}_{\mathcal{I}_1,\mathcal{I}_2})$, Theorem \ref{th:shared_convergence} and Theorem \ref{th:better_than_not_modul} to the Smoothed Split Conformal framework}} 

The functions $ \bar{s}^{c}_{\mathcal{I}_1}$ and $ \bar{s}_{\mathcal{I}_1}$ are defined as in the Split Conformal framework, except for: $k$ ($\gamma$ respectively) that is the $\lceil l + \tau_{n+1} - (l+1) \alpha  \rceil$th ($\lceil m + \tau_{n+1} - (m+1) \alpha  \rceil$th respectively) smallest value in the corresponding set; similarly to the Split Conformal framework, if $\lceil m + \tau_{n+1} - (m+1) \alpha \rceil > m$ then $\mathcal{H}_1=\mathcal{I}_1$ and if $\lceil m + \tau_{n+1} - (m+1) \alpha \rceil \leq 0$ we arbitrarily set $\bar{s}_{j,\mathcal{I}_1}=s^0_j$. Theorem \ref{th:shared_convergence} and Theorem \ref{th:better_than_not_modul} still hold by substituting $\lceil (l+1)(1-\alpha) \rceil, \lceil (m+1)(1-\alpha) \rceil$ with $\lceil l + \tau_{n+1} - (l+1) \alpha  \rceil, \lceil m + \tau_{n+1} - (m+1) \alpha  \rceil$.

\subsection*{B Supplementary Figures}
\label{sec::sup_fig}

\renewcommand{\thefigure}{B.1} 
\begin{figure}[!h]
\begin{center}
\includegraphics[width=10cm,height=7.7308cm]{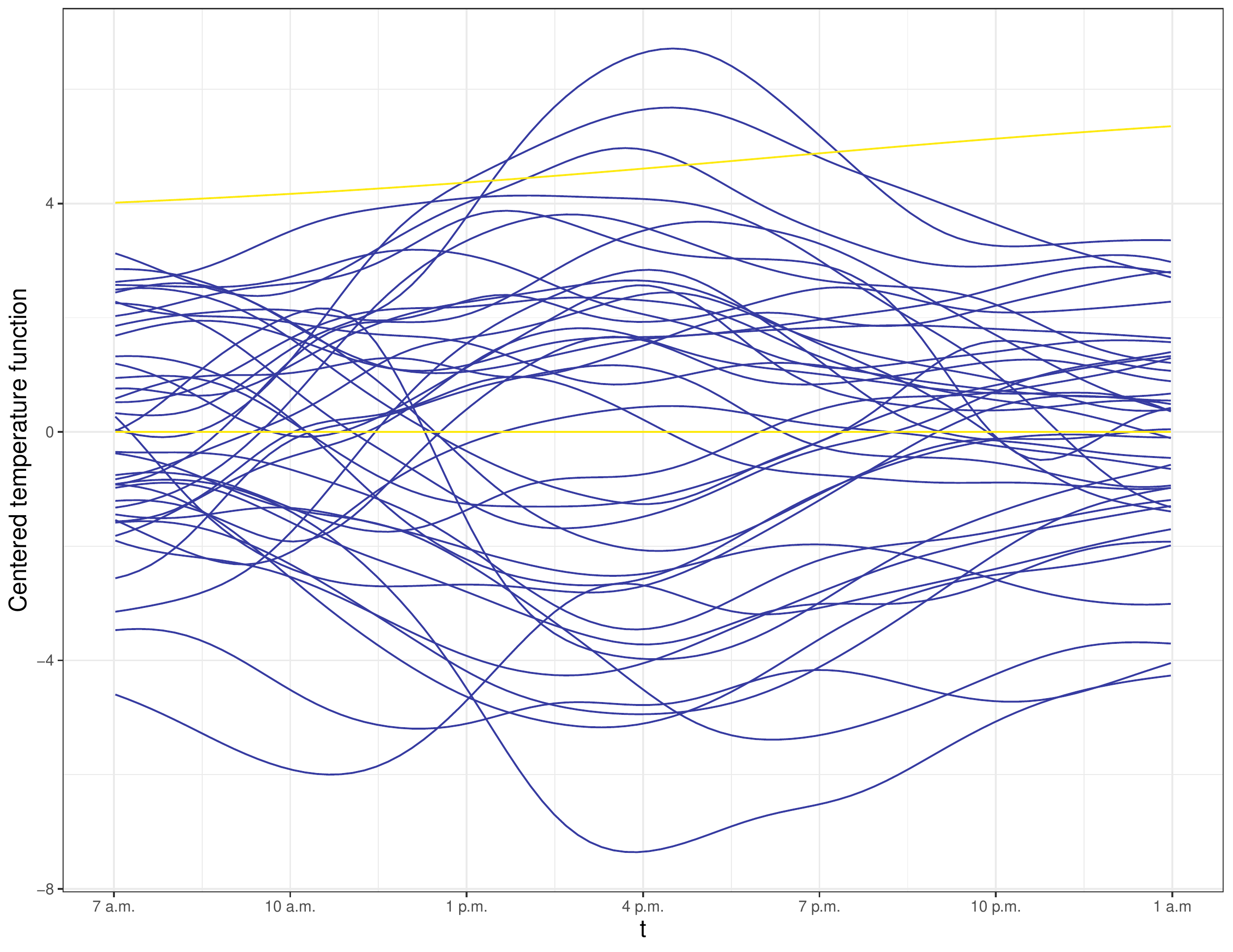} 
\end{center}
\caption{Temperature function (after subtracting the average daily temperature function of the period considered) in degrees Celsius for the observed days (blue curves) and for the two hypothetical days (yellow curves). \label{fig::cov_bikemi_matsup} }
\end{figure}

\section*{Acknowledgements}

Prof. Vantini and Dr. Fontana acknowledge the financial support from Accordo Quadro ASI-POLIMI ``Attivit\`a di Ricerca e Innovazione'' n. 2018-5-HH.0, collaboration agreement between the Italian Space Agency and Politecnico di Milano. The authors are deeply grateful to Clear Channel Italia S.p.A, which provided the data for the case study, and to Agostino Torti for providing part of the code used in the case study.

\bibliographystyle{agsm}

\end{document}